\documentclass[english,aps,nofootinbib,longbibliography,notitlepage,superscriptaddress,prd,preprintnumbers,showkeys,twocolumn]{revtex4-1}
\pdfoutput=1

\usepackage[usenames,dvipsnames,svgnames,table]{xcolor}
\usepackage{verbatim}
\usepackage[T1]{fontenc}
\usepackage[latin9]{inputenc}
\usepackage{babel}
\usepackage{float}
\usepackage{graphicx}
\usepackage{svg}
\usepackage{hyperref}
\usepackage{amsthm,amsmath,amssymb,amsfonts}
\usepackage[mathscr]{eucal}
\usepackage{microtype}
\usepackage{esint}
\usepackage{amsbsy}
\usepackage{color}
\usepackage[capitalise]{cleveref}
\usepackage{breakurl}
\usepackage[normalem]{ulem}
\usepackage{bbm}
\usepackage{mathtools}
\usepackage{braket}
\usepackage{comment}
\usepackage[super]{nth}
\usepackage{csquotes}
\usepackage{lmodern}
\usepackage{hyphenat}

\definecolor{lightgray}{gray}{0.9}
\definecolor{Amber}{rgb}{1.0, 0.75, 0.0}
\definecolor{blizzardblue}{rgb}{0.67, 0.9, 0.93}
\definecolor{myblue}{HTML}{0273B2}
\colorlet{myblueLight}{myblue!35}

\usepackage{tikz}
\usepackage{pgfplots}
\pgfplotsset{compat=newest,every axis plot/.append style={line width=1pt}}

\allowdisplaybreaks

\makeatletter

\pdfpageheight\paperheight
\pdfpagewidth\paperwidth

\@ifundefined{textcolor}{}{%
 \definecolor{BLACK}{gray}{0}
 \definecolor{WHITE}{gray}{1}
 \definecolor{RED}{rgb}{1,0,0}
 \definecolor{GREEN}{rgb}{0,1,0}
 \definecolor{BLUE}{rgb}{0,0,1}
 \definecolor{CYAN}{cmyk}{1,0,0,0}
 \definecolor{MAGENTA}{cmyk}{0,1,0,0}
 \definecolor{YELLOW}{cmyk}{0,0,1,0}
}

\DeclareMathOperator\erf{erf}

\makeatother

\makeatletter

\DeclareRobustCommand{\rcite}[1]{%
  \rcite@aux#1,\@nil{#1}%
}
\def\rcite@aux#1,#2\@nil#3{%
  \if\relax#2\relax
    Ref.~\cite{#3}%
  \else
    Refs.~\cite{#3}%
  \fi
}
\makeatother
\definecolor{wine}{RGB}{136,34,85}
\definecolor{teal}{RGB}{0,85,102}
\hypersetup{
    colorlinks = true,
    citecolor = {wine},
    linkcolor = {teal},
    urlcolor = {teal},
}

\newcommand{\be}{\begin{equation}}
\newcommand{\ee}{\end{equation}}
\newcommand{\ba}{\begin{eqnarray}}
\newcommand{\ea}{\end{eqnarray}}

\newcommand{\dd}{\mathrm{d}}

\begin{document}

\preprint{IFT-UAM/CSIC-25-36}

\title{Has DESI detected exponential quintessence?}

\author{Yashar Akrami}
\email{yashar.akrami@csic.es}
\affiliation{Instituto de F\'isica Te\'orica (IFT) UAM-CSIC, C/ Nicol\'as Cabrera 13-15, Campus de Cantoblanco UAM, 28049 Madrid, Spain}
\affiliation{CERCA/ISO, Department of Physics, Case Western Reserve University, 10900 Euclid Avenue, Cleveland, OH 44106, USA}
\affiliation{Astrophysics Group \& Imperial Centre for Inference and Cosmology, Department of Physics, Imperial College London, Blackett Laboratory, Prince Consort Road, London SW7 2AZ, United Kingdom}

\author{George Alestas}
\email{savvas.nesseris@csic.es}
\affiliation{Instituto de F\'isica Te\'orica (IFT) UAM-CSIC, C/ Nicol\'as Cabrera 13-15, Campus de Cantoblanco UAM, 28049 Madrid, Spain}

\author{Savvas Nesseris}
\email{g.alestas@csic.es}
\affiliation{Instituto de F\'isica Te\'orica (IFT) UAM-CSIC, C/ Nicol\'as Cabrera 13-15, Campus de Cantoblanco UAM, 28049 Madrid, Spain}

\date{\today}

\begin{abstract}
The new Dark Energy Spectroscopic Instrument (DESI) DR2 results have strengthened the possibility that dark energy is dynamical, i.e., it has evolved over the history of the Universe. One simple, but theoretically well motivated and widely studied, physical model of dynamical dark energy is minimally coupled, single-field quintessence $\phi$ with an exponential potential $V(\phi)=V_0\,e^{-\lambda\phi}$. We perform a full Bayesian statistical analysis of the model using the DESI DR2 data, in combination with other cosmological observations, to constrain the model's parameters and to compare its goodness of fit to that of the standard $\Lambda$CDM model. We find that the quintessence model provides a significantly better fit to the data, both when the spatial curvature of the Universe is fixed to zero and when it is allowed to vary. The significance of the preference varies between $\sim3.3\sigma$ and $\sim3.8\sigma$, depending on whether the curvature density parameter $\Omega_K$ is fixed or varied. We obtain the values $0.698^{+0.173}_{-0.202}$  and $0.722^{+0.182}_{-0.208}$ at the $68.3\%$ (i.e., $1\sigma$) confidence level for the parameter $\lambda$ in the absence and presence of $\Omega_K$, respectively, which imply $\sim3.5\sigma$ preference for a nonzero $\lambda$. We also obtain $\Omega_K=0.003\pm 0.001$, which implies $\sim3\sigma$ preference for a positive $\Omega_K$, i.e., a negative curvature. Finally, we discuss the differences between quintessence and phenomenological parametrizations of the dark energy equation-of-state parameter, in particular the Chevallier-Polarski-Linder (CPL) parametrization, as well as a few caveats to our results.
\end{abstract}

\keywords{cosmic acceleration, dark energy, quintessence, Dark Energy Spectroscopic Instrument (DESI)}
\preprint{}
\maketitle


\section{Introduction}
\label{sec:Introduction}
The recent Dark Energy Spectroscopic Instrument (DESI) baryon acoustic oscillations (BAO) results \cite{DESI:2025zgx,Lodha:2025qbg} once again confirmed the nearly-three-decade observations \cite{SupernovaCosmologyProject:1998vns,SupernovaSearchTeam:1998fmf} that the Universe is currently accelerating. This is independent of how we parametrize the mechanism behind the cosmic acceleration, whether it is the cosmological constant $\Lambda$ of the standard cosmological $\Lambda$CDM model, one or more dynamical fields that we collectively call dark energy \cite{Copeland:2006wr}, some modification of Einstein's theory of general relativity on cosmological scales \cite{Clifton:2011jh,Bamba:2012cp,Joyce:2014kja,Koyama:2015vza,Bull:2015stt,Ishak:2018his,Ferreira:2019xrr,CANTATA:2021asi}, or something that does not require new physics. This is, however, not the primary reason why one should be excited about the DESI results. As an update on DESI's initial findings \cite{DESI:2024mwx}, the new results provide the first hints of potential deviations from a cosmological constant, which, if confirmed with $>5\sigma$ significance, will falsify the $\Lambda$CDM model.

It has become common for almost all observational collaborations to report their constraints on deviations from $\Lambda$ through the so-called Chevallier-Polarski-Linder (CPL) parametrization \cite{Chevallier:2000qy,Linder:2002et}
\begin{equation}
    \label{eqn:CPLasTaylorina}
    w_\mathrm{DE}(a) = w_0 + w_a(1-a)\,,
\end{equation}
where $a$ is the scale factor of the Friedmann-Lema\^{i}tre-Robertson-Walker (FLRW) metric (with $a$ being set to 1 at the present time) and $w_0$ and $w_a$ are constants; see \rcite{Lodha:2025qbg,Kessler:2025kju,Shlivko:2025fgv} for recently proposed alternative parametrizations. Here, $w_\mathrm{DE}$ is the equation of state parameter for dark energy,
\begin{equation}
    w_\mathrm{DE} = \frac{P_\mathrm {DE}}{\rho_\mathrm{DE}}\,,
\end{equation}
with $P_\mathrm {DE}$ the pressure and $\rho_\mathrm{DE}$ the energy density of dark energy. In terms of the CPL phenomenological parametrization \eqref{eqn:CPLasTaylorina}, the DESI results, in combination with different supernova datasets, imply a preference for dynamical dark energy over $\Lambda$ that ranges from $2.8\sigma$ to $4.2\sigma$ depending on the supernova dataset \cite{DESI:2025zgx,Lodha:2025qbg}. Additionally, the best-fit CPL model, if taken at face value, implies that dark energy has crossed the phantom divide at a redshift $z_\mathrm{PD}\sim0.4$, i.e., $w_\mathrm{DE}>-1$ at $z<z_\mathrm{PD}$ and $w_\mathrm{DE}<-1$ at $z>z_\mathrm{PD}$. This then implies that simple, minimally coupled, scalar-field models of dark energy, or quintessence \cite{Copeland:2006wr}, may be on the verge of being excluded observationally, since $w_\mathrm{DE}$ is always larger than $-1$ in those models. 

As some of us argued in \rcite{Nesseris:2025lke}, the CPL parametrization is, however, not a physical model based on physical theories and is only the truncation to first order of a Taylor expansion of $w_\mathrm{DE}(a)$ around its current value. Unless observations tell us that higher-order terms in the expansion are nearly zero, we may, by choosing the CPL parametrization, be injecting an extraordinary amount of information into the data that is not necessarily consistent with the observations. \rcite{Nesseris:2025lke} shows that this is indeed the case for DESI, as the DESI constraints on the two CPL parameters $w_0$ and $w_a$ are already significantly weaker when we add one or two higher-order Taylor terms and marginalize over the extra parameters, allowing a wide range of dark energy models to be consistent with the data even if they do not feature any phantom-like behavior implied by the CPL parametrization. \rcite{Nesseris:2025lke} then suggested that one should either test the cosmological constant $\Lambda$ in a frequentist way using data-derived statistics---such as principal components \cite{Huterer:2002hy}---or directly compare physical models---as opposed to low-dimensional phenomenological parametrizations---to $\Lambda$ in a Bayesian way. Motivated by these, we place, in this paper, state-of-the-art constraints on arguably the simplest model of dynamical dark energy, i.e., minimally coupled, single-field quintessence with an exponential potential.

Not only are quintessence models interesting from the point of view of cosmology---for providing a simple mechanism for dynamical dark energy---they are also one of the focuses of low-energy phenomenological studies of quantum gravity theories, in particular string theory; see, e.g., \rcite{Agrawal:2018own,Cicoli:2018kdo,Akrami:2018ylq,Han:2018yrk,Raveri:2018ddi,Akrami:2020zfz,Andriot:2024jsh,Bhattacharya:2024hep,Andriot:2024sif,Alestas:2024gxe,Bhattacharya:2024kxp,Andriot:2025gyr}. A universe filled with a  canonical scalar field $\phi$ minimally coupled to gravity as dark energy, nonrelativistic matter (composed of baryonic and cold dark matter), and radiation is given by the action
\begin{equation}
S = \int \dd^4 x \sqrt{-g} \left[ \frac{1}{2}M_\mathrm{Pl}^2 R - \frac{1}{2} g^{\mu \nu} \partial_{\mu}\phi \partial_{\nu}\phi -V(\phi) + \mathcal{L}_{\rm{m,r}} \right]\,,
\end{equation}
where $M_\mathrm{Pl}$ is the reduced Plank mass, $R$ is the Ricci scalar, and $\mathcal{L}_{\rm{m,r}}$ is the matter and radiation Lagrangian density. The metric $g_{\mu\nu}$ is assumed to be of an FLRW form describing a homogeneous, isotropic, negatively curved (``hyperbolic'') universe;
a homogeneous, isotropic, flat (``Euclidean'') universe; and
a homogeneous, isotropic, positively curved (``spherical'') universe. $V(\phi)$ is the potential term for the scalar field, the simplest form of which is, arguably, the exponential function
\begin{equation}
V(\phi)=V_0\,e^{-\lambda\phi}\,,\label{expoform}
\end{equation}
where $\lambda$ is a free parameter. This simple potential is particularly motivated by the fact that every time one computes a scalar-field potential in ultraviolet-complete theories of quantum gravity such as string theory, it is a sum of exponential functions, and in all known examples ever checked, one obtains only effective single-field models with exponential scalar potentials. It is particularly important to measure the value of $\lambda$ from observations, as $\lambda\geq\sqrt{2}$ in all known examples of string theory scenarios in the so-called asymptotic regions of moduli space, where the theory is under perturbative control \cite{Maldacena:2000mw,Hertzberg:2007wc,Obied:2018sgi,Andriot:2019wrs,Andriot:2020lea,Calderon-Infante:2022nxb,Shiu:2023fhb,Shiu:2023nph,Cremonini:2023suw,Hebecker:2023qke,VanRiet:2023cca,Seo:2024fki}---this has been conjectured to hold universally \cite{Rudelius:2021azq}. A number of studies estimated $\lambda$ using the cosmological observations, both prior to \cite{Agrawal:2018own,Akrami:2018ylq,Raveri:2018ddi,Akrami:2020zfz} and following \cite{Bhattacharya:2024hep,Ramadan:2024kmn,Alestas:2024gxe} the DESI DR1 results \cite{DESI:2024mwx}.

In this paper, we perform a full Bayesian statistical analysis of  the quintessence model of cosmic acceleration with the exponential potential \eqref{expoform} and provide updated constraints on its parameters, in particular the parameter $\lambda$, using the recent DESI DR2 data. We study the model both when spatial flatness is imposed and when the spatial curvature density parameter $\Omega_K$ is allowed to vary, place constraints on relevant parameters, and test the model in terms of goodness-of-fit criteria when compared to the standard $\Lambda$CDM model. This will allow us to know whether this particular model of dynamical dark energy is favored over $\Lambda$CDM, and if it is, at what confidence level. We also compare our findings to those from fitting phenomenological parametrizations such as CPL and discuss the implications of the comparison.

The paper is organized as follows. In \cref{sec:datamethods}, we describe the models and the parameterizations studied in the paper, the observational data, and the statistical techniques and quantities used in our analysis. In \cref{sec:results}, we present the constraints we obtain on the free parameters fo the models and parametrizations, and compare their goodness of fit through a Bayesian model selection method. We also discuss the significance of our results and their implications for models of dark energy, in particular in relation to phenomenological parametrizations. We conclude in \cref{sec:conclusions} and provide additional details in \cref{app:A,app:B}.

\section{Data and method}
\label{sec:datamethods}
We use the DESI DR2 BAO data \cite{DESI:2025zgx, Lodha:2025qbg} in combination with the {\it Planck} cosmic microwave background (CMB) distance priors \cite{Planck:2018vyg} and the Dark Energy Survey Year 5 (DESY5) type Ia supernova data \cite{DES:2024jxu}. The CMB distance priors are comprised of the acoustic scale---measuring the CMB temperature in the transverse direction, the CMB shift parameter---measuring the peak spacing of the CMB temperature in the power spectrum, and the baryon density $\omega_\mathrm{b}\equiv\Omega_\mathrm{b}h^{2}$. Here, $\Omega_\mathrm{b}$ is the baryon density parameter and $h\equiv H_0/(100\,\mathrm{km}\,\mathrm{s}^{-1}\,\mathrm{Mpc}^{-1})$, where $H_0$ is the present value of the Hubble expansion rate $H$; see \rcite{Zhai:2018vmm} for details. In order to be consistent with \rcite{DESI:2024mwx}, we also assume the approximate expression of the drag-epoch sound horizon of \rcite{Brieden:2022heh} and a Big Bang nucleosynthesis prior on the baryon density $\Omega_\mathrm{b}h^2=0.02218 \pm 0.00055$ \cite{Schoneberg:2024ifp}. This means that we will have $N=1845$ data points in total.

\begin{table*}
\renewcommand{\arraystretch}{1.3}
\rowcolors{1}{}{lightgray}
    \centering
    \setlength\tabcolsep{0pt}
    \begin{tabular}{ |c|c|c|c|c|c| }
    \hline
    \rowcolor{myblueLight}
    ~Parameter/Quantity~ &  ~~Flat $\Lambda$CDM~~ & ~~$\Lambda$CDM$+\Omega_{K}$~~ & ~~Flat $\phi$CDM~~ & ~~$\phi$CDM$+\Omega_{K}$~~ & ~~~Flat CPL~~~  \\
    \hline
    $\Omega_{\rm{m}}$ & ~~$0.305\pm 0.003$~~ & $0.306\pm 0.003$ & ~~$0.315\pm 0.005$~~ & $0.316\pm 0.006$ & $0.320\pm 0.006$ \\
    $\Omega_{K}$ & $\cdots$ &$0.003\pm 0.001$ & $\cdots$ & $0.003\pm 0.001$ & $\cdots$\\
    $H_0$& $67.96\pm 0.23$ & $68.48\pm 0.30$ & $66.81\pm 0.56$ & $67.29\pm 0.62$ &$66.73\pm 0.57$\\
    $\lambda$ &$\cdots$&$\cdots$ & $0.698^{+0.173}_{-0.202}$  & $0.722^{+0.182}_{-0.208}$ &$\cdots$\\
    $V_0$ &$\cdots$&$\cdots$ & $2.207\pm 0.389$  & $2.299\pm 0.332$ & $\cdots$\\
    $w_0$ & $-1$ & $-1$& $\cdots$& $\cdots$& ~ $-0.751\pm 0.058$\,\\
    $w_a$ & 0 & 0& $\cdots$& $\cdots$& ~$-0.877\pm 0.231$\\
    $\chi^2$ & $1680.70$ & $1672.08$ & $1673.98$ & $1664.11$ & 1660.65\\
    $\Delta\ln B$ & 0 & $-1.55$ & 4.03 & 3.55 & 6.84 \\
    \hline
    \end{tabular}
    \caption{The $1\sigma$ (i.e., $68.3\%$ confidence level) constraints on model parameters obtained through MCMC scans of the parameter space for the models and parametrizations studied in the present work, along with their corresponding minimum $\chi^2$ and $\Delta\ln B$, where $\Delta\ln B_X = \ln B_X- \ln B_{\mathrm{flat} \Lambda \mathrm{CDM}}$ for a given model or parametrization $X$. $\Delta\ln B_X>0$ implies evidence in favor of model $X$ over flat $\Lambda$CDM, while $\Delta\ln B_X<0$ implies evidence in favor of flat $\Lambda$CDM over model $X$.
    These can be interpreted through the updated Jeffreys' scale: $\Delta\ln B_{X}<1.1$ implies that model $X$ is comparable with flat $\Lambda$CDM, with neither one being distinctly preferred to the other; $1.1<\Delta \ln B_{X}<3$ implies weak evidence favoring model $X$ to flat $\Lambda$CDM; $3<\Delta \ln B_{X}<5$ implies moderate evidence favoring model $X$ to flat $\Lambda$CDM; $\Delta \ln B_{X,Y}>5$ implies strong support for model $X$ over flat $\Lambda$CDM. We use $N=1845$ data points in our statistical analysis.}
    \label{tab:MCMC_bf}
\end{table*}

With these, we perform a Markov chain Monte Carlo (MCMC) analysis of flat $\Lambda\mathrm{CDM}$, $\Lambda\mathrm{CDM}\!+\!\Omega_K$, flat quintessence $\phi$CDM, and the quintessence model with free $\Omega_K$, $\phi \mathrm{CDM}\!+\!\Omega_K$, as well as the CPL parametrization of the dark energy equation-of-state parameter $w_\mathrm{DE}(z)$ with $\Omega_K=0$, which we simply call flat CPL, in order to compare the goodness of fit of these models and parametrizations and to place constraints on their free parameters.\footnote{We do not use \emph{shooting} algorithms to solve the equations for our quintessence cases as they would significantly slow down the numerical computations. In our statistical analysis, the present values of some of the parameters, such as $H_0$, $\Omega_\mathrm{m}$, and $\Omega_K$, are ``derived parameters'' and not direct outputs of the MCMC scans. Therefore, we need to appropriately convert the resulting constraints on the varied parameters into the ones on the derived parameters through a postprocessing step---this choice of numerical technique does not affect the final results.} In each case, we generate at least $\sim$150\,000 points and ensure that the chains are well converged. 

Finally, we use the \texttt{MCEvidence} code \cite{Heavens:2017afc} to compute the Bayes factors $B_X$ of $\Lambda\mathrm{CDM}\!+\!\Omega_K$, flat $\phi$CDM, $\phi \mathrm{CDM}\!+\!\Omega_K$, and flat CPL relative to flat $\Lambda$CDM, in order to assess their quality of fit; see also \cref{app:A}. The results can be interpreted through the Jeffreys' scale \cite{Jeffreys:1939xee, Trotta:2008qt} and the updated Jeffreys' scale \cite{John:2002gg,Nesseris:2012cq}, which are empirically calibrated with thresholds corresponding to the odds at specific ratios \cite{Trotta:2008qt}. Based on the updated scale, if we form the differences $\Delta\ln B_{X,Y} \equiv \ln B_X-\ln B_Y $ for models $X$ and $Y$, then $\Delta\ln B_{X,Y}<1.1$ implies that the two models are comparable, with neither one being distinctly preferred to the other. When $1.1<\Delta \ln B_{X,Y}<3$, there is weak evidence favoring model $X$ to model $Y$. For $3<\Delta \ln B_{X,Y}<5$, the evidence is moderate, and if $\Delta \ln B_{X,Y}>5$, there is strong support for model $X$ over model $Y$.

\section{Results and discussion}
\label{sec:results}
\begin{figure}[!t]
\centering
\includegraphics[width=0.48\textwidth]{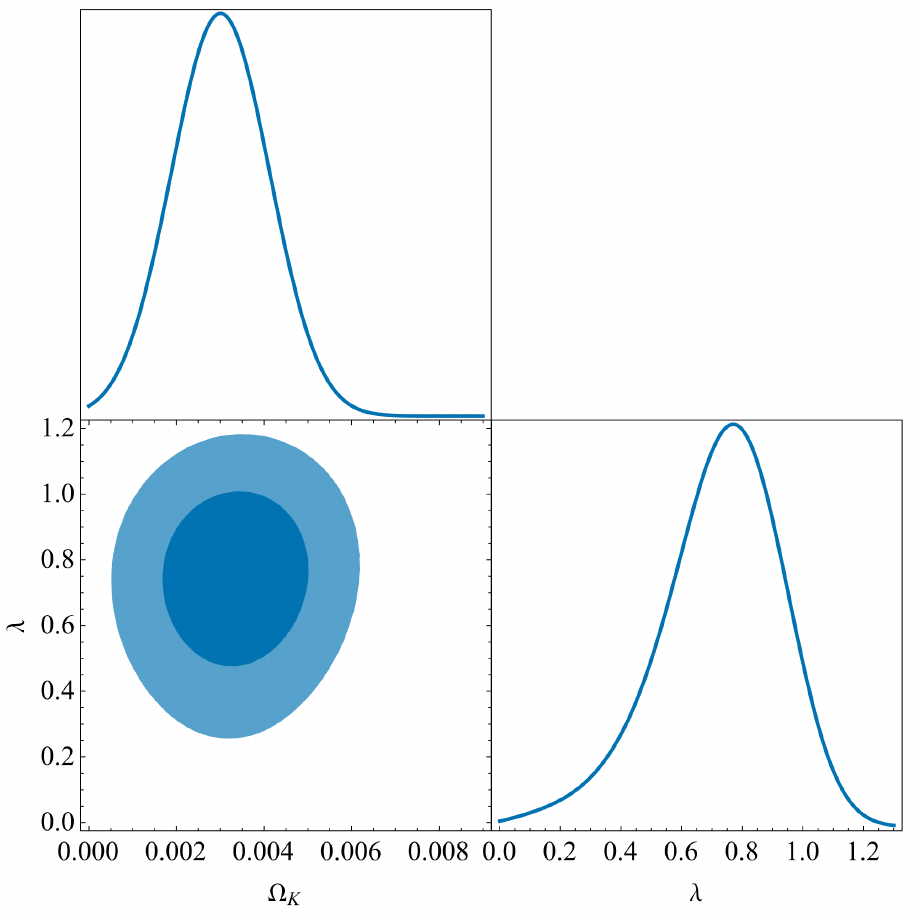}
\caption{A triangle plot showing the two-dimensional, marginalized $\lambda$-$\Omega_K$ $68.3\%$ and $95.5\%$ confidence error contours (lower left panel) and the one-dimensional, marginalized posterior probability distribution functions for $\Omega_K$ (upper panel) and $\lambda$ (lower right panel) for the quintessence model of dark energy with an exponential potential, $V(\phi)=V_0\,e^{-\lambda\phi}$, when the spatial curvature density parameter $\Omega_{K}$ is allowed to vary as a free parameter with a uniform prior. }
\label{fig:cont_l_Om}
\end{figure}

\begin{figure}[!h]
\centering
\includegraphics[width=0.48\textwidth]{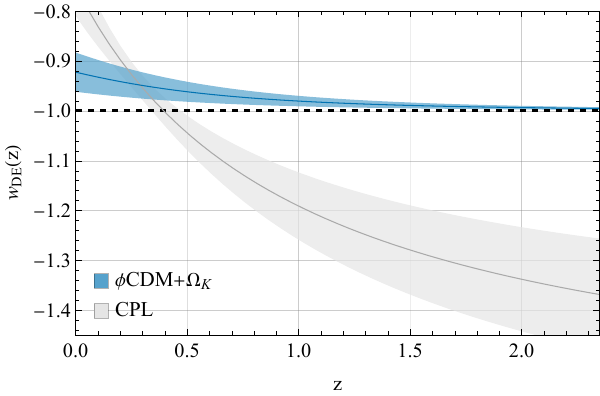}
\caption{The redshift evolution of the dark energy equation-of-state parameter corresponding to the Bayesian mean quintessence model with an exponential potential when the spatial curvature density parameter $\Omega_{K}$ is allowed to vary as a free parameter (blue solid curve). For comparison, we have also shown the best-fit CPL parametrization with $\Omega_{K}=0$ (gray solid curve). The color-shaded regions show the corresponding 1$\sigma$ (i.e., $68.3\%$ confidence level) error regions.}
\label{fig:plot_wDE} 
\end{figure}

The main results of our MCMC analysis are presented in \cref{tab:MCMC_bf}, as well as in \cref{fig:cont_l_Om,fig:plot_wDE}. In Table~\ref{tab:MCMC_bf}, we provide $1\sigma$ (i.e., $68.3\%$ confidence level) constraints on the free parameters obtained through the MCMC exploration of the parameter space for each model or parametrization---$\Omega_\mathrm{m}$ is the present value of the matter density parameter. We also provide the values of the minimum $\chi^2$ and $\Delta\ln B_X$ corresponding to each model, where, again, $\Delta\ln B_X = \ln B_X- \ln B_{\mathrm{flat} \Lambda \mathrm{CDM}}$. This means that $\Delta\ln B_X>0$ implies evidence in favor of model $X$ over flat $\Lambda$CDM, while $\Delta\ln B_X<0$ implies evidence in favor of flat $\Lambda$CDM over model $X$. Our results have several implications.

We first notice that $\Lambda$CDM in the presence of free $\Omega_K$ is weakly disfavored compared to flat $\Lambda$CDM, according to the Jeffreys' scale, even though it offers a lower minimum $\chi^2$. The exponential quintessence model, on the other had, is moderately favored over flat $\Lambda$CDM (with $\Delta \ln B\sim 4$, corresponding to $\sim3.3\sigma$ significance,\footnote{See \cref{app:B} for details.} for flat $\phi \mathrm{CDM}$ and with $\Delta \ln B\sim 3.5$, corresponding to $\sim3.2\sigma$ significance, for $\phi \mathrm{CDM}\!+\!\Omega_K$), while it is strongly favored over $\Lambda\mathrm{CDM}\!+\!\Omega_K$ (with $\Delta \ln B\sim 5.6$, corresponding to $\sim3.8\sigma$ significance, for flat $\phi \mathrm{CDM}$ and with $\Delta \ln B\sim 5.1$, corresponding to $\sim3.6\sigma$ significance, for $\phi \mathrm{CDM}\!+\!\Omega_K$). 
Comparing the results for flat CPL to the ones for other cases shows, however, that flat CPL provides a better fit to the data (with $\Delta \ln B\sim 6.8$, corresponding to $\sim4.1\sigma$ significance, compared to flat $\Lambda\mathrm{CDM}$ and with $\Delta \ln B\sim 8.4$, corresponding to $\sim4.5\sigma$ significance, compared to $\Lambda\mathrm{CDM}\!+\!\Omega_K$), but it is important to note that CPL is only a phenomenological parametrization of the dark energy equation-of-state parameter, while quintessence is a physical model. 

In any Bayesian model selection, one should, in principle, also take into account the theoretical priors on the models. Even though these are difficult to quantify, if not impossible, one should be very careful when drawing conclusions about the models---this effect of priors would almost certainly (and potentially strongly) weaken any preference for a phenomenological parametrization like CPL over a theoretically motivated model like quintessence.

In terms of the constraints on free parameters, the most interesting ones are the limits on the parameter $\lambda$ for quintessence and on the spatial curvature density parameter $\Omega_K$ for both $\Lambda$CDM and quintessence. Our results show that the data prefer a nonzero value for $\lambda$ at the $\sim99.95\%$ confidence level (CL), corresponding to $\sim3.5\sigma$ significance, implying that quintessence is preferred to $\Lambda$CDM both with and without varying $\Omega_K$---this is consistent with our previous discussions of the Bayes factor quantification of goodness of fit and model selection. Additionally, our results show that there is an about $3\sigma$ preference for a positive $\Omega_K$ in both $\Lambda$CDM and quintessence scenarios. In \cref{fig:cont_l_Om}, we provide the $68.3\%$ and $95.5\%$ two-dimensional confidence contours and one-dimensional marginalized probability density functions for the quintessence model when the spatial curvature parameter $\Omega_{K}$ is allowed to vary as a free parameter with a uniform prior. We also notice from \cref{tab:MCMC_bf} that the preferred value of the Hubble constant $H_0$ varies slightly from model to model. It is particularly interesting to note that the data prefers a lower value of $H_0$ for flat $\phi \mathrm{CDM}$ compared to flat $\Lambda$CDM, as expected for quintessence; it is also lower than the value for $\phi \mathrm{CDM}\!+\!\Omega_K$, which is lower than the one for $\Lambda\mathrm{CDM}\!+\!\Omega_K$.

In \cref{fig:plot_wDE}, we show the evolution of the quintessence equation-of-state parameter $w_\mathrm{\phi}(z)$ as a function of redshift $z$ for the (Bayesian) mean $\phi \mathrm{CDM}\!+\!\Omega_K$ model (blue solid curve) versus the mean $w_\mathrm{CPL}(z)$ for the CPL parametrization (gray solid curve). The shaded regions show the corresponding $68.3\%$ confidence regions, estimated via standard error propagation and using the covariance matrices obtained from the MCMC analysis. As expected, $w_\mathrm{\phi}(z)$ remains above $-1$ at all times, while $w_\mathrm{CPL}(z)$ crosses the phantom line at $z\sim 0.4$. Since quintessence provides a good fit to the data, the figure clearly shows that the recent DESI measurements do not necessarily imply phantom crossing for dark energy. Note, again, that even though the CPL parametrization is statistically preferred to the quintessence model, the preference is not very significant, and additionally, CPL is not a physical model---it is only a simple, two-parameter, phenomenological parametrization of the dark energy equation-of-state parameter $w_\mathrm{DE}(z)$. Our results, therefore, demonstrate that it is too early to conclude that the DESI data (in combination with other observations)---if remains significant in the future---imply the need for dark energy to cross the phantom line.

\begin{figure*}[!t]
\centering
\includegraphics[width=0.49\textwidth]{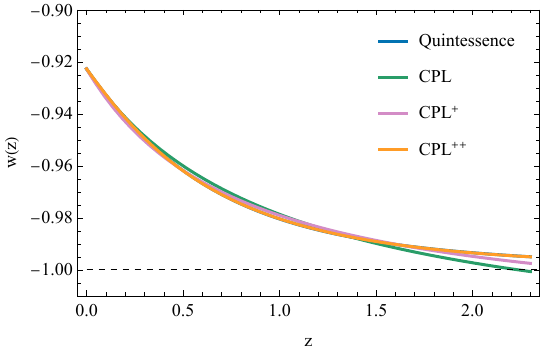}
\includegraphics[width=0.48\textwidth]{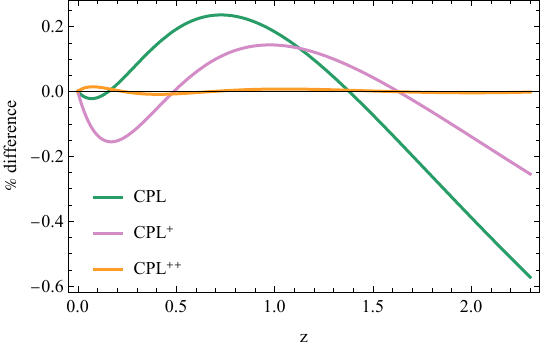}
\includegraphics[width=0.49\textwidth]{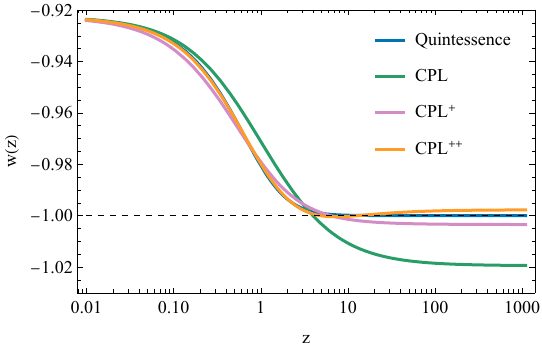}
\includegraphics[width=0.48\textwidth]{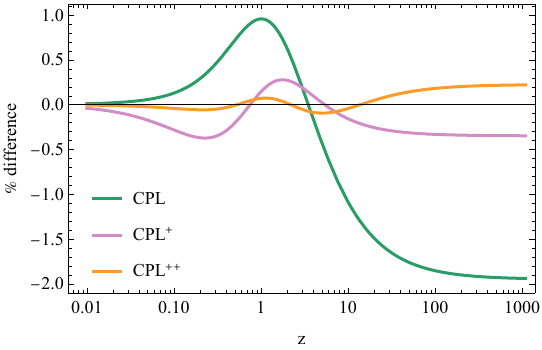}
\caption{\label{fig:plot_wDE_CPL} Right panels: the Bayesian mean quintessence equation-of-state parameter $w_\phi(z)$ in comparison to the equation-of-state parameters $w_{\mathrm{CPL}}(z)$, $w_{\mathrm{CPL}^{+}}(z)$, and $w_{\mathrm{CPL}^{++}}(z)$ for the CPL, CPL$^+$, and CPL$^{++}$ parametrizations, respectively, best fit to the $w_\phi(z)$ over a small range of redshift, $z\in[0,2.35]$ (upper panels), and a large range of redshift, $z\in[0,1100]$ (lower panels). Left panels: percentage differences between the $w_{\mathrm{CPL}}(z)$, $w_{\mathrm{CPL}^{+}}(z)$, and $w_{\mathrm{CPL}^{++}}(z)$ of the right panels and the mean $w_\phi(z)$ over the same ranges of redshift.
}
\end{figure*}

As an additional investigation, it is interesting to know how well the CPL parametrization can mimic the quintessence equation-of-state parameter over different ranges of redshift. This will shed light on why CPL provides a better fit to the data. For that reason, we compute the $w_\mathrm{CPL}(z)$ that follows our mean $w_\mathrm{\phi}(z)$ as closely as possible, with the condition that $w_0$ exactly matches $w_\mathrm{\phi}(z=0)$. In order to ensure that CPL is the truncated, first-order Taylor expansion of $w_\mathrm{\phi}(z)$ around $a=1$, we compute the fitted $w_\mathrm{CPL}(z)$ in terms of $a$. The mean $w_\mathrm{\phi}(z)$ and its corresponding $w_\mathrm{CPL}(z)$ are shown in Fig.~\ref{fig:plot_wDE_CPL}, where the upper panels are for the CPL fit to quintessence over a small range of redshift, $z\in[0,2.35]$, and the lower panels are for the CPL fit over a large range of redshift, $z\in[0,1100]$. The figure shows that CPL is able to provide a $\lesssim2\%$ fit to quintessence at all redshifts, all the way to the redshift of recombination, even though the best-fit $w_\mathrm{CPL}(z)$ goes slightly below $-1$ at high redshifts, which is forbidden by quintessence \cite{Nesseris:2006er}. This is a very good fit given the sensitivity of current and upcoming cosmological surveys. The question then is, why is the quintessence's fit to the data not as good as the CPL's fit if CPL provides such a good approximation to the quintessence equation-of-state parameter? Comparing \cref{fig:plot_wDE_CPL} with \cref{fig:plot_wDE} may suggest that the reason is in the high-redshift behavior of the $w_\mathrm{CPL}(z)$ that is best fit to the data, as the data seems to require phantom crossing with values of $w_\mathrm{DE}(z)$ significantly (i.e., with $>3\sigma$) less than $-1$ while the CPL fit to quintessence does not provide that behavior.

In order to know whether this is in fact the reason, we also compute two higher-order extensions of CPL that best fit the quintessence equation-of-state parameter over the same redshift ranges, again with their $w_0$ fixed to $w_\mathrm{\phi}(z=0)$. These are the CPL$^+$ and CPL$^{++}$ parametrizations introduced in \rcite{Nesseris:2025lke}. By extending the CPL equation-of-state parameter to
\begin{equation}
    \label{eqn:CPL++}
    w_\mathrm{DE}(a) = w_0 + w_a(1-a) + w_b(1-a)^2 + w_c(1-a)^3\,,
\end{equation}
CPL$^+$ corresponds to varying $w_0$, $w_a$, and $w_b$ while fixing $w_c=0$, and CPL$^{++}$ corresponds to varying all the parameters $w_0$, $w_a$, $w_b$, and $w_c$. In \cref{fig:plot_wDE_CPL}, we show the CPL$^{+}$ and CPL$^{++}$ parametrizations that best fit our $w_\mathrm{\phi}(z)$. The figure shows that these simple extensions---especially CPL$^{++}$---provide better fits to quintessence, especially when the large range of redshift is considered---this is expected, as discussed in \rcite{Nesseris:2025lke}. 

These effects can be seen more clearly by computing the values of the additional parameters for the CPL$^{+}$ and CPL$^{++}$ parametrizations that best fit the quintessence model. We obtain $(w_a, w_b)=(-0.126, 0.027)$ and $(w_a, w_b,w_c)=(-0.101,-0.090,0.125)$ for CPL$^{+}$ and CPL$^{++}$, respectively, when fitted over the low range of redshift, and $(w_a, w_b)=(-0.145,0.064)$ and $(w_a,w_b,w_c)=(-0.110,-0.050,0.086)$, respectively, when fitted over the large range of redshift. The fact that CPL$^{+}$ and CPL$^{++}$ require nonzero values for the additional parameters is the reason why they provide better fits to quintessence, but what is interesting is that now the error regions around the CPL$^{+}$ and CPL$^{++}$ $w_\mathrm{DE}(z)$ best fit to the data increase significantly and include almost entirely the mean non-phantom $w_\mathrm{\phi}(z)$---cf. Fig. 2 in \rcite{Nesseris:2025lke}. These again tell us that the results of dark energy searches using simple Taylor expansions should be interpreted with great care, as these parametrizations may easily miss physical models and lead to incorrect conclusions.

\section{Conclusions}
\label{sec:conclusions}
In this work, we have performed a full Bayesian statistical analysis of minimally coupled, single-field quintessence with an exponential potential as a theoretically highly motivated and arguably the simplest model of dynamical dark energy. We have used the latest results of the DESI large-scale structure survey in combination with the $Planck$ microwave background distance priors and the DESY5 type Ia supernova data. We have assessed the quality of fit of the quintessence model compared to the standard $\Lambda$CDM model and have estimated its free parameters.

We have found that the exponential quintessence model is significantly preferred by the data to $\Lambda$CDM regardless of whether we vary the spatial curvature density parameter $\Omega_K$ as a free parameter or fix it to zero, i.e., assuming a flat universe. We have estimated the significance for this preference to be between $\sim3.3\sigma$ and $\sim3.8\sigma$, corresponding to a difference in the log Bayes factors of $\Delta \ln B\sim 4$ to $\Delta \ln B \sim 5.6$, depending on whether or not $\Omega_K$ is varied. These results imply moderate to strong evidence for quintessence compared to $\Lambda$CDM. In terms of the constraints on parameters, we have estimated $\sim 3.5\sigma$ and $\sim 3\sigma$ preferences for nonzero values of the quintessence parameter $\lambda$ and of the curvature density parameter $\Omega_K$, respectively.

We have also compared $\Lambda$CDM and the quintessence model to the flat CPL parameterization of the dark energy equation-of-state parameter. We have found a $\Delta \ln B$ of $\sim 6.8$ for flat CPL compared to flat $\Lambda$CDM, implying strong evidence, but only a $\Delta \ln B$ of $\sim 2.8$ when compared to the flat quintessence model, implying weak evidence. One should also note that CPL is only a phenomenological parametrization of the dark energy equation-of-state parameter $w_\mathrm{DE}(z)$, while quintessence is a well motivated physical model, with a (potentially much stronger) theoretical prior compared to CPL---this must be taken into account in any Bayesian comparison of models.

In order to better understand the reason behind the small observational preference for CPL compared to quintessence, we have also investigated the ability of the CPL parametrization to mimic the quintessence equation-of-state parameter over small and large ranges of redshift, when the present value of $w_\mathrm{CPL}(z)$, i.e., $w_0$, is fixed to the present value of $w_\mathrm{\phi}(z)$, i.e., when $w_0 = w_\phi(z=0)$---we have assumed this due to the fact that the CPL parametrization is the first-order Taylor expansion of $w_\mathrm{DE}(1-a)$, valid at low redshifts. We have found that the CPL best fit to the quintessence $w_\phi(z)$ exhibits phantom behavior at high redshifts, even though this behavior is forbidden by quintessence \cite{Nesseris:2006er}. An additional investigation of higher-order extensions of the CPL parametrization has shown that they provide better fits to the quintessence equation-of-state parameter while simultaneously exhibiting larger error regions around their best-fit equation-of-state parameters that include non-phantom models like quintessence. Our analysis of the exponential quintessence model, therefore, explicitly confirms the concern raised in \rcite{Nesseris:2025lke} about using low-dimensional phenomenological parameterizations like CPL. Our results, along with those of \rcite{Nesseris:2025lke}, demonstrate that dark energy searches using simple Taylor expansions, and in general, phenomenological parametrizations, should be interpreted with great care, as they tend to inject information into the data that is not necessarily consistent with the observations, and not necessarily demanded by physical models.

Finally, it is important to note that there are caveats to our conclusions about the evidence for exponential quintessence and the significance of that evidence. In our analysis, we have taken the DESI BAO data and error bars at face value, as provided by the collaboration, while concerns have been raised about possible outliers in the data \cite{Sapone:2024ltl,Colgain:2024xqj,Chudaykin:2024gol} and using the standard BAO analysis techniques, particularly for testing models beyond the standard flat $\Lambda$CDM model; see, e.g., \rcite{Anselmi:2018vjz,Anselmi:2022exn}. Additionally, we should note that all of the $\chi^2$ values in \cref{tab:MCMC_bf} are significantly lower than the number of degrees of freedom ($=1845$) minus the number of parameters, resulting in reduced $\chi^2$ values that are $<1$ in all the cases we have studied---this is possibly due to overestimation of the supernovae errors; see \rcite{Nielsen:2015pga, Sah:2024csa}. All these caveats should be considered when interpreting our results and conclusions.

\begin{table*}
\small
\renewcommand{\arraystretch}{1.3}
\rowcolors{1}{}{lightgray}
    \centering
    \setlength\tabcolsep{0pt}
    \begin{tabular}{ |c|c|c|c|c|}
    \hline
    \rowcolor{myblueLight}
    ~Model/Parametrization~ & ~~AIC~~ & ~~$\Delta$AIC~~ & ~~BIC~~ & ~~$\Delta$BIC~~ \\
    \hline     
     Flat $\Lambda$CDM & ~~$1684.70$~~ & $0$ & ~~$1695.74$~~ & $0$ \\
     ~~$\Lambda$CDM$+\Omega_{K}$~~ & $1678.08$ & $6.62$ & ~~$1694.64$~~ & $~~1.10$ \\
     Flat $\phi$CDM & $1681.98$ & $2.72$ & ~~$1704.06$~~ & $-8.32$ \\
     $\phi$CDM$+\Omega_{K}$& $1674.11$ & $10.59$ & ~~$1701.71$~~ & $-5.97$ \\
     Flat CPL & $1668.65$  & $16.05$ & ~~$1690.73$~~ & $~~5.01$ \\
    \hline
    \end{tabular}
    \caption{Values of the Akaike information criterion (AIC) and the Bayesian information criterion (BIC) for the models and parametrizations considered in the present work, along with their differences, $\Delta$AIC and $\Delta$BIC, compared to the ones computed for flat $\Lambda$CDM; cf. \cref{tab:MCMC_bf}.}
    \label{tab:bic_aic}
\end{table*}

\acknowledgments{
Y.A. acknowledges support by the Spanish Research Agency (Agencia Estatal de Investigaci\'on)'s grant RYC2020-030193-I/AEI/10.13039/501100011033, by the European Social Fund (Fondo Social Europeo) through the  Ram\'{o}n y Cajal program within the State Plan for Scientific and Technical Research and Innovation (Plan Estatal de Investigaci\'on Cient\'ifica y T\'ecnica y de Innovaci\'on) 2017-2020, by the Spanish Research Agency through the grant IFT Centro de Excelencia Severo Ochoa No CEX2020-001007-S funded by MCIN/AEI/10.13039/501100011033, and by the Spanish National Research Council (CSIC) through the Talent Attraction grant 20225AT025.
S.N. acknowledges support from the research project PID2021-123012NB-C43 and the Spanish Research Agency (Agencia Estatal de Investigaci\'on) through the grant IFT Centro de Excelencia Severo Ochoa No CEX2020-001007-S, funded by MCIN/AEI/10.13039/501100011033. G.A. is supported by the Spanish Attraccion de Talento Contract No. 2019-T1/TIC-13177 granted by the Comunidad de Madrid. We thank the IFT pizza club for helpful discussions. We also thank Stefano Anselmi, Eoin \'{O} Colg\'{a}in, and Glenn D. Starkman for useful comments on a previous version of the manuscript.} This work made use of the IFT Hydra cluster.

\appendix
\section{Information criteria} \label{app:A}
In the Bayesian model comparison framework, one must consider not only the goodness of fit for competing models, as measured by the corresponding minimum $\chi^2$ values, but also the number of free parameters in each model. The more free parameters a model possesses, the more it should be penalized. Although the most proper way of performing the Bayesian model selection procedure is by computing the Bayes factors for the models compared to each other, as we have done in \cref{sec:results}, here we discuss two other approaches to model selection and the caution one should exercise when using them.

These are the co-called Akaike information criterion (AIC) \cite{AkaikeCrit}, defined as
\be
\mathrm{AIC} \equiv -2 \, \ln \mathcal{L}_\textrm{max}+2n=\chi_\textrm{min}^2 +2n \,, \label{eq:AIC}
\ee
and the Bayesian information criterion (BIC) \cite{Schwarz1978}, defined as
\be
\mathrm{BIC} \equiv \chi_\textrm{min}^2 + n\ln N \,, \label{eq:BIC}
\ee
where $\mathcal{L}_\textrm{max}$ is the maximum likelihood, $\chi_\textrm{min}$ is the minimum $\chi^2$, $n$ is the total number of free parameters of the considered model, and $N$ is the total number of data points used in the statistical analysis. 

Using the definitions \eqref{eq:AIC} and \eqref{eq:BIC}, we compute the AIC and BIC values for all the models and parametrizations considered in the present work. We provide in \cref{tab:bic_aic} the differences $\Delta \mathrm{AIC} \equiv \mathrm{AIC}_\textrm{flat $\Lambda$CDM} - \mathrm{AIC}_X$ and $\Delta \mathrm{BIC} \equiv \mathrm{BIC}_\textrm{flat $\Lambda$CDM} - \mathrm{BIC}_X$ for each model or parametrization $X$.

The results of \cref{tab:bic_aic} illustrate a number of issues. We first notice that the $\Delta \mathrm{AIC}$ and $\Delta \mathrm{BIC}$ values do not show a trend similar to the one we obtained in \cref{sec:results} using the Bayes factor computations. Even though both $\Delta \mathrm{AIC}$ and $\Delta \mathrm{BIC}$ show the strongest preference for flat CPL, similarly to the $\Delta\ln B$ values of \cref{sec:results}, their values for $\Lambda$CDM and quintessence do not necessarily imply the same conclusions as those obtained by the $\Delta\ln B$ analysis.

The $\Delta \mathrm{AIC}$ values show  that the quintessence model when the curvature parameter $\Omega_K$ is allowed to vary, i.e., $\phi \mathrm{CDM}\!+\!\Omega_K$, is very strongly favored over flat $\Lambda$CDM and moderately favored over $\Lambda\mathrm{CDM}\!+\!\Omega_K$, which does not agree with our results of \cref{tab:MCMC_bf} showing moderate preference for $\phi \mathrm{CDM}\!+\!\Omega_K$ over flat $\Lambda$CDM and strong preference over $\Lambda\mathrm{CDM}\!+\!\Omega_K$. Additionally, our $\Delta\ln B$ analysis of \cref{tab:MCMC_bf} shows that flat $\Lambda$CDM is weakly favored compared to $\Lambda\mathrm{CDM}\!+\!\Omega_K$, while the $\Delta \mathrm{AIC}$ results imply that flat $\Lambda$CDM is strongly disfavored compared to $\Lambda\mathrm{CDM}\!+\!\Omega_K$. \cref{tab:MCMC_bf} also tells us that flat $\phi \mathrm{CDM}$ and $\phi \mathrm{CDM}\!+\!\Omega_K$ are comparable, with a slight preference for flat $\phi \mathrm{CDM}$, while \cref{tab:bic_aic} implies that flat $\phi \mathrm{CDM}$ is strongly disfavored compared to $\phi \mathrm{CDM}\!+\!\Omega_K$.

The $\Delta \mathrm{BIC}$ results are even more confusing. Now both flat $\phi \mathrm{CDM}$ and $\phi \mathrm{CDM}\!+\!\Omega_K$ are found to be strongly disfavored compared to flat $\Lambda\mathrm{CDM}$,  $\Lambda\mathrm{CDM}\!+\!\Omega_K$, and flat CPL, which is in complete disagreement with the $\Delta\ln B$ and $\Delta \mathrm{AIC}$ findings. 

We should, however, note that both $\Delta \mathrm{AIC}$ and $\Delta \mathrm{BIC}$ are only computationally fast approximations to statistically proper measures such as the Bayes factor, and if these proper statistical measures are available, one should use them instead of the approximations which may not always agree with them---that is why we only report our $\Delta\ln B$ results in \cref{sec:results} and use them to draw conclusions. The reason why the $\Delta \mathrm{AIC}$ values agree better with the $\Delta\ln B$ values compared to the $\Delta \mathrm{BIC}$ values is that BIC penalizes a model with a higher number of free parameters much more strongly, as the term $2n$ in AIC is replaced by the term $n\ln N$ in BIC, which becomes significantly larger by increasing the number of data points $N$.

\section{Conversion of Bayes factors to frequentist significance ($p$-values)}
\label{app:B}
Here we briefly summarize the procedure we followed in \cref{sec:results} to translate the Bayes factors we obtained for different models, through a Bayesian model comparison analysis, into frequentist significance quantities (i.e., $p$-values and number of sigmas).

Let us assume a complex model $X$ and a simpler model $Y$ for which the Bayes factor is $B_{XY}= 1/B_{YX}$. The lower bound on the $p$-value $\wp$, defined as the probability that a test statistic would be as large as or larger than the observed value assuming the null hypothesis (i.e., the simpler model) is true, can be determined by \cite{Sellke01022001,Trotta:2008qt}
\be 
B_{XY}\le -\frac1{\mathrm{e}\, \wp\,\ln \wp}\,,\label{eq:p-val}
\ee 
where $\mathrm{e}\approx2.71828$ is Euler's number. Given a value for $B_{XY}$ or $\Delta \ln B_{XY}$, such as the ones given in Table~\ref{tab:MCMC_bf}, we can invert \eqref{eq:p-val} to find the corresponding $p$-value in terms of the Lambert $W$ function $W_k(z)$, for some integer $k$, by taking the correct branch ($k=-1$). Then, the $p$-value can be translated into the frequentist number of sigmas by assuming a normal distribution via the usual expression $\sigma = \sqrt{2}\, \erf^{-1}\left(1-\wp\right)$, where $ \erf^{-1}(z)$ is the inverse error function.\footnote{The Lambert $W$ function $W_k(z)$ and the inverse error function $ \erf^{-1}(z)$ can be calculated, to arbitrary precision, with \texttt{Mathematica} and via commands \texttt{ProductLog[k,z]} and \texttt{InverseErf[z]}, respectively.}

\bibliography{bibliography}

\begin{thebibliography}{67}%
\makeatletter
\providecommand \@ifxundefined [1]{%
 \@ifx{#1\undefined}
}%
\providecommand \@ifnum [1]{%
 \ifnum #1\expandafter \@firstoftwo
 \else \expandafter \@secondoftwo
 \fi
}%
\providecommand \@ifx [1]{%
 \ifx #1\expandafter \@firstoftwo
 \else \expandafter \@secondoftwo
 \fi
}%
\providecommand \natexlab [1]{#1}%
\providecommand \enquote  [1]{``#1''}%
\providecommand \bibnamefont  [1]{#1}%
\providecommand \bibfnamefont [1]{#1}%
\providecommand \citenamefont [1]{#1}%
\providecommand \href@noop [0]{\@secondoftwo}%
\providecommand \href [0]{\begingroup \@sanitize@url \@href}%
\providecommand \@href[1]{\@@startlink{#1}\@@href}%
\providecommand \@@href[1]{\endgroup#1\@@endlink}%
\providecommand \@sanitize@url [0]{\catcode `\\12\catcode `\$12\catcode `\&12\catcode `\#12\catcode `\^12\catcode `\_12\catcode `\%12\relax}%
\providecommand \@@startlink[1]{}%
\providecommand \@@endlink[0]{}%
\providecommand \url  [0]{\begingroup\@sanitize@url \@url }%
\providecommand \@url [1]{\endgroup\@href {#1}{\urlprefix }}%
\providecommand \urlprefix  [0]{URL }%
\providecommand \Eprint [0]{\href }%
\providecommand \doibase [0]{http://dx.doi.org/}%
\providecommand \selectlanguage [0]{\@gobble}%
\providecommand \bibinfo  [0]{\@secondoftwo}%
\providecommand \bibfield  [0]{\@secondoftwo}%
\providecommand \translation [1]{[#1]}%
\providecommand \BibitemOpen [0]{}%
\providecommand \bibitemStop [0]{}%
\providecommand \bibitemNoStop [0]{.\EOS\space}%
\providecommand \EOS [0]{\spacefactor3000\relax}%
\providecommand \BibitemShut  [1]{\csname bibitem#1\endcsname}%
\let\auto@bib@innerbib\@empty
\bibitem [{\citenamefont {Abdul~Karim}\ \emph {et~al.}(2025)\citenamefont {Abdul~Karim} \emph {et~al.}}]{DESI:2025zgx}%
  \BibitemOpen
  \bibfield  {author} {\bibinfo {author} {\bibfnamefont {M.}~\bibnamefont {Abdul~Karim}} \emph {et~al.} (\bibinfo {collaboration} {DESI}),\ }\bibfield  {title} {\enquote {\bibinfo {title} {{DESI DR2 Results II: Measurements of Baryon Acoustic Oscillations and Cosmological Constraints}},}\ }\href@noop {} {\  (\bibinfo {year} {2025})},\ \Eprint {http://arxiv.org/abs/2503.14738} {arXiv:2503.14738 [astro-ph.CO]} \BibitemShut {NoStop}%
\bibitem [{\citenamefont {Lodha}\ \emph {et~al.}(2025)\citenamefont {Lodha} \emph {et~al.}}]{Lodha:2025qbg}%
  \BibitemOpen
  \bibfield  {author} {\bibinfo {author} {\bibfnamefont {K.}~\bibnamefont {Lodha}} \emph {et~al.},\ }\bibfield  {title} {\enquote {\bibinfo {title} {{Extended Dark Energy analysis using DESI DR2 BAO measurements}},}\ }\href@noop {} {\  (\bibinfo {year} {2025})},\ \Eprint {http://arxiv.org/abs/2503.14743} {arXiv:2503.14743 [astro-ph.CO]} \BibitemShut {NoStop}%
\bibitem [{\citenamefont {Perlmutter}\ \emph {et~al.}(1999)\citenamefont {Perlmutter} \emph {et~al.}}]{SupernovaCosmologyProject:1998vns}%
  \BibitemOpen
  \bibfield  {author} {\bibinfo {author} {\bibfnamefont {S.}~\bibnamefont {Perlmutter}} \emph {et~al.} (\bibinfo {collaboration} {Supernova Cosmology Project}),\ }\bibfield  {title} {\enquote {\bibinfo {title} {{Measurements of $\Omega$ and $\Lambda$ from 42 high redshift supernovae}},}\ }\href {\doibase 10.1086/307221} {\bibfield  {journal} {\bibinfo  {journal} {Astrophys. J.}\ }\textbf {\bibinfo {volume} {517}},\ \bibinfo {pages} {565--586} (\bibinfo {year} {1999})},\ \Eprint {http://arxiv.org/abs/astro-ph/9812133} {arXiv:astro-ph/9812133} \BibitemShut {NoStop}%
\bibitem [{\citenamefont {Riess}\ \emph {et~al.}(1998)\citenamefont {Riess} \emph {et~al.}}]{SupernovaSearchTeam:1998fmf}%
  \BibitemOpen
  \bibfield  {author} {\bibinfo {author} {\bibfnamefont {Adam~G.}\ \bibnamefont {Riess}} \emph {et~al.} (\bibinfo {collaboration} {Supernova Search Team}),\ }\bibfield  {title} {\enquote {\bibinfo {title} {{Observational evidence from supernovae for an accelerating universe and a cosmological constant}},}\ }\href {\doibase 10.1086/300499} {\bibfield  {journal} {\bibinfo  {journal} {Astron. J.}\ }\textbf {\bibinfo {volume} {116}},\ \bibinfo {pages} {1009--1038} (\bibinfo {year} {1998})},\ \Eprint {http://arxiv.org/abs/astro-ph/9805201} {arXiv:astro-ph/9805201} \BibitemShut {NoStop}%
\bibitem [{\citenamefont {Copeland}\ \emph {et~al.}(2006)\citenamefont {Copeland}, \citenamefont {Sami},\ and\ \citenamefont {Tsujikawa}}]{Copeland:2006wr}%
  \BibitemOpen
  \bibfield  {author} {\bibinfo {author} {\bibfnamefont {Edmund~J.}\ \bibnamefont {Copeland}}, \bibinfo {author} {\bibfnamefont {M.}~\bibnamefont {Sami}}, \ and\ \bibinfo {author} {\bibfnamefont {Shinji}\ \bibnamefont {Tsujikawa}},\ }\bibfield  {title} {\enquote {\bibinfo {title} {{Dynamics of dark energy}},}\ }\href {\doibase 10.1142/S021827180600942X} {\bibfield  {journal} {\bibinfo  {journal} {Int. J. Mod. Phys. D}\ }\textbf {\bibinfo {volume} {15}},\ \bibinfo {pages} {1753--1936} (\bibinfo {year} {2006})},\ \Eprint {http://arxiv.org/abs/hep-th/0603057} {arXiv:hep-th/0603057} \BibitemShut {NoStop}%
\bibitem [{\citenamefont {Clifton}\ \emph {et~al.}(2012)\citenamefont {Clifton}, \citenamefont {Ferreira}, \citenamefont {Padilla},\ and\ \citenamefont {Skordis}}]{Clifton:2011jh}%
  \BibitemOpen
  \bibfield  {author} {\bibinfo {author} {\bibfnamefont {Timothy}\ \bibnamefont {Clifton}}, \bibinfo {author} {\bibfnamefont {Pedro~G.}\ \bibnamefont {Ferreira}}, \bibinfo {author} {\bibfnamefont {Antonio}\ \bibnamefont {Padilla}}, \ and\ \bibinfo {author} {\bibfnamefont {Constantinos}\ \bibnamefont {Skordis}},\ }\bibfield  {title} {\enquote {\bibinfo {title} {{Modified Gravity and Cosmology}},}\ }\href {\doibase 10.1016/j.physrep.2012.01.001} {\bibfield  {journal} {\bibinfo  {journal} {Phys. Rept.}\ }\textbf {\bibinfo {volume} {513}},\ \bibinfo {pages} {1--189} (\bibinfo {year} {2012})},\ \Eprint {http://arxiv.org/abs/1106.2476} {arXiv:1106.2476 [astro-ph.CO]} \BibitemShut {NoStop}%
\bibitem [{\citenamefont {Bamba}\ \emph {et~al.}(2012)\citenamefont {Bamba}, \citenamefont {Capozziello}, \citenamefont {Nojiri},\ and\ \citenamefont {Odintsov}}]{Bamba:2012cp}%
  \BibitemOpen
  \bibfield  {author} {\bibinfo {author} {\bibfnamefont {Kazuharu}\ \bibnamefont {Bamba}}, \bibinfo {author} {\bibfnamefont {Salvatore}\ \bibnamefont {Capozziello}}, \bibinfo {author} {\bibfnamefont {Shin'ichi}\ \bibnamefont {Nojiri}}, \ and\ \bibinfo {author} {\bibfnamefont {Sergei~D.}\ \bibnamefont {Odintsov}},\ }\bibfield  {title} {\enquote {\bibinfo {title} {{Dark energy cosmology: the equivalent description via different theoretical models and cosmography tests}},}\ }\href {\doibase 10.1007/s10509-012-1181-8} {\bibfield  {journal} {\bibinfo  {journal} {Astrophys. Space Sci.}\ }\textbf {\bibinfo {volume} {342}},\ \bibinfo {pages} {155--228} (\bibinfo {year} {2012})},\ \Eprint {http://arxiv.org/abs/1205.3421} {arXiv:1205.3421 [gr-qc]} \BibitemShut {NoStop}%
\bibitem [{\citenamefont {Joyce}\ \emph {et~al.}(2015)\citenamefont {Joyce}, \citenamefont {Jain}, \citenamefont {Khoury},\ and\ \citenamefont {Trodden}}]{Joyce:2014kja}%
  \BibitemOpen
  \bibfield  {author} {\bibinfo {author} {\bibfnamefont {Austin}\ \bibnamefont {Joyce}}, \bibinfo {author} {\bibfnamefont {Bhuvnesh}\ \bibnamefont {Jain}}, \bibinfo {author} {\bibfnamefont {Justin}\ \bibnamefont {Khoury}}, \ and\ \bibinfo {author} {\bibfnamefont {Mark}\ \bibnamefont {Trodden}},\ }\bibfield  {title} {\enquote {\bibinfo {title} {{Beyond the Cosmological Standard Model}},}\ }\href {\doibase 10.1016/j.physrep.2014.12.002} {\bibfield  {journal} {\bibinfo  {journal} {Phys. Rept.}\ }\textbf {\bibinfo {volume} {568}},\ \bibinfo {pages} {1--98} (\bibinfo {year} {2015})},\ \Eprint {http://arxiv.org/abs/1407.0059} {arXiv:1407.0059 [astro-ph.CO]} \BibitemShut {NoStop}%
\bibitem [{\citenamefont {Koyama}(2016)}]{Koyama:2015vza}%
  \BibitemOpen
  \bibfield  {author} {\bibinfo {author} {\bibfnamefont {Kazuya}\ \bibnamefont {Koyama}},\ }\bibfield  {title} {\enquote {\bibinfo {title} {{Cosmological Tests of Modified Gravity}},}\ }\href {\doibase 10.1088/0034-4885/79/4/046902} {\bibfield  {journal} {\bibinfo  {journal} {Rept. Prog. Phys.}\ }\textbf {\bibinfo {volume} {79}},\ \bibinfo {pages} {046902} (\bibinfo {year} {2016})},\ \Eprint {http://arxiv.org/abs/1504.04623} {arXiv:1504.04623 [astro-ph.CO]} \BibitemShut {NoStop}%
\bibitem [{\citenamefont {Bull}\ \emph {et~al.}(2016)\citenamefont {Bull}, \citenamefont {Akrami} \emph {et~al.}}]{Bull:2015stt}%
  \BibitemOpen
  \bibfield  {author} {\bibinfo {author} {\bibfnamefont {Philip}\ \bibnamefont {Bull}}, \bibinfo {author} {\bibfnamefont {Yashar}\ \bibnamefont {Akrami}},  \emph {et~al.},\ }\bibfield  {title} {\enquote {\bibinfo {title} {{Beyond $\Lambda$CDM: Problems, solutions, and the road ahead}},}\ }\href {\doibase 10.1016/j.dark.2016.02.001} {\bibfield  {journal} {\bibinfo  {journal} {Phys. Dark Univ.}\ }\textbf {\bibinfo {volume} {12}},\ \bibinfo {pages} {56--99} (\bibinfo {year} {2016})},\ \Eprint {http://arxiv.org/abs/1512.05356} {arXiv:1512.05356 [astro-ph.CO]} \BibitemShut {NoStop}%
\bibitem [{\citenamefont {Ishak}(2019)}]{Ishak:2018his}%
  \BibitemOpen
  \bibfield  {author} {\bibinfo {author} {\bibfnamefont {Mustapha}\ \bibnamefont {Ishak}},\ }\bibfield  {title} {\enquote {\bibinfo {title} {{Testing General Relativity in Cosmology}},}\ }\href {\doibase 10.1007/s41114-018-0017-4} {\bibfield  {journal} {\bibinfo  {journal} {Living Rev. Rel.}\ }\textbf {\bibinfo {volume} {22}},\ \bibinfo {pages} {1} (\bibinfo {year} {2019})},\ \Eprint {http://arxiv.org/abs/1806.10122} {arXiv:1806.10122 [astro-ph.CO]} \BibitemShut {NoStop}%
\bibitem [{\citenamefont {Ferreira}(2019)}]{Ferreira:2019xrr}%
  \BibitemOpen
  \bibfield  {author} {\bibinfo {author} {\bibfnamefont {Pedro~G.}\ \bibnamefont {Ferreira}},\ }\bibfield  {title} {\enquote {\bibinfo {title} {{Cosmological Tests of Gravity}},}\ }\href {\doibase 10.1146/annurev-astro-091918-104423} {\bibfield  {journal} {\bibinfo  {journal} {Ann. Rev. Astron. Astrophys.}\ }\textbf {\bibinfo {volume} {57}},\ \bibinfo {pages} {335--374} (\bibinfo {year} {2019})},\ \Eprint {http://arxiv.org/abs/1902.10503} {arXiv:1902.10503 [astro-ph.CO]} \BibitemShut {NoStop}%
\bibitem [{\citenamefont {Akrami}\ \emph {et~al.}(2021{\natexlab{a}})\citenamefont {Akrami} \emph {et~al.}}]{CANTATA:2021asi}%
  \BibitemOpen
  \bibfield  {author} {\bibinfo {author} {\bibfnamefont {Yashar}\ \bibnamefont {Akrami}} \emph {et~al.} (\bibinfo {collaboration} {CANTATA}),\ }\href {\doibase 10.1007/978-3-030-83715-0} {\emph {\bibinfo {title} {{Modified Gravity and Cosmology. An Update by the CANTATA Network}}}},\ edited by\ \bibinfo {editor} {\bibfnamefont {Emmanuel~N.}\ \bibnamefont {Saridakis}}, \bibinfo {editor} {\bibfnamefont {Ruth}\ \bibnamefont {Lazkoz}}, \bibinfo {editor} {\bibfnamefont {Vincenzo}\ \bibnamefont {Salzano}}, \bibinfo {editor} {\bibfnamefont {Paulo}\ \bibnamefont {Vargas~Moniz}}, \bibinfo {editor} {\bibfnamefont {Salvatore}\ \bibnamefont {Capozziello}}, \bibinfo {editor} {\bibfnamefont {Jose}\ \bibnamefont {Beltr\'an~Jim\'enez}}, \bibinfo {editor} {\bibfnamefont {Mariafelicia}\ \bibnamefont {De~Laurentis}}, \ and\ \bibinfo {editor} {\bibfnamefont {Gonzalo~J.}\ \bibnamefont {Olmo}}\ (\bibinfo  {publisher} {Springer},\ \bibinfo {year} {2021})\ \Eprint {http://arxiv.org/abs/2105.12582} {arXiv:2105.12582 [gr-qc]}
  \BibitemShut {NoStop}%
\bibitem [{\citenamefont {Adame}\ \emph {et~al.}(2025)\citenamefont {Adame} \emph {et~al.}}]{DESI:2024mwx}%
  \BibitemOpen
  \bibfield  {author} {\bibinfo {author} {\bibfnamefont {A.~G.}\ \bibnamefont {Adame}} \emph {et~al.} (\bibinfo {collaboration} {DESI}),\ }\bibfield  {title} {\enquote {\bibinfo {title} {{DESI 2024 VI: cosmological constraints from the measurements of baryon acoustic oscillations}},}\ }\href {\doibase 10.1088/1475-7516/2025/02/021} {\bibfield  {journal} {\bibinfo  {journal} {JCAP}\ }\textbf {\bibinfo {volume} {02}},\ \bibinfo {pages} {021} (\bibinfo {year} {2025})},\ \Eprint {http://arxiv.org/abs/2404.03002} {arXiv:2404.03002 [astro-ph.CO]} \BibitemShut {NoStop}%
\bibitem [{\citenamefont {Chevallier}\ and\ \citenamefont {Polarski}(2001)}]{Chevallier:2000qy}%
  \BibitemOpen
  \bibfield  {author} {\bibinfo {author} {\bibfnamefont {Michel}\ \bibnamefont {Chevallier}}\ and\ \bibinfo {author} {\bibfnamefont {David}\ \bibnamefont {Polarski}},\ }\bibfield  {title} {\enquote {\bibinfo {title} {{Accelerating universes with scaling dark matter}},}\ }\href {\doibase 10.1142/S0218271801000822} {\bibfield  {journal} {\bibinfo  {journal} {Int. J. Mod. Phys. D}\ }\textbf {\bibinfo {volume} {10}},\ \bibinfo {pages} {213--224} (\bibinfo {year} {2001})},\ \Eprint {http://arxiv.org/abs/gr-qc/0009008} {arXiv:gr-qc/0009008} \BibitemShut {NoStop}%
\bibitem [{\citenamefont {Linder}(2003)}]{Linder:2002et}%
  \BibitemOpen
  \bibfield  {author} {\bibinfo {author} {\bibfnamefont {Eric~V.}\ \bibnamefont {Linder}},\ }\bibfield  {title} {\enquote {\bibinfo {title} {{Exploring the expansion history of the universe}},}\ }\href {\doibase 10.1103/PhysRevLett.90.091301} {\bibfield  {journal} {\bibinfo  {journal} {Phys. Rev. Lett.}\ }\textbf {\bibinfo {volume} {90}},\ \bibinfo {pages} {091301} (\bibinfo {year} {2003})},\ \Eprint {http://arxiv.org/abs/astro-ph/0208512} {arXiv:astro-ph/0208512} \BibitemShut {NoStop}%
\bibitem [{\citenamefont {Kessler}\ \emph {et~al.}(2025)\citenamefont {Kessler}, \citenamefont {Escamilla}, \citenamefont {Pan},\ and\ \citenamefont {Di~Valentino}}]{Kessler:2025kju}%
  \BibitemOpen
  \bibfield  {author} {\bibinfo {author} {\bibfnamefont {Daniel~A.}\ \bibnamefont {Kessler}}, \bibinfo {author} {\bibfnamefont {Luis~A.}\ \bibnamefont {Escamilla}}, \bibinfo {author} {\bibfnamefont {Supriya}\ \bibnamefont {Pan}}, \ and\ \bibinfo {author} {\bibfnamefont {Eleonora}\ \bibnamefont {Di~Valentino}},\ }\bibfield  {title} {\enquote {\bibinfo {title} {{One-parameter dynamical dark energy: Hints for oscillations}},}\ }\href@noop {} {\  (\bibinfo {year} {2025})},\ \Eprint {http://arxiv.org/abs/2504.00776} {arXiv:2504.00776 [astro-ph.CO]} \BibitemShut {NoStop}%
\bibitem [{\citenamefont {Shlivko}\ \emph {et~al.}(2025)\citenamefont {Shlivko}, \citenamefont {Steinhardt},\ and\ \citenamefont {Steinhardt}}]{Shlivko:2025fgv}%
  \BibitemOpen
  \bibfield  {author} {\bibinfo {author} {\bibfnamefont {David}\ \bibnamefont {Shlivko}}, \bibinfo {author} {\bibfnamefont {Paul~J.}\ \bibnamefont {Steinhardt}}, \ and\ \bibinfo {author} {\bibfnamefont {Charles~L.}\ \bibnamefont {Steinhardt}},\ }\bibfield  {title} {\enquote {\bibinfo {title} {{Optimal parameterizations for observational constraints on thawing dark energy}},}\ }\href@noop {} {\  (\bibinfo {year} {2025})},\ \Eprint {http://arxiv.org/abs/2504.02028} {arXiv:2504.02028 [astro-ph.CO]} \BibitemShut {NoStop}%
\bibitem [{\citenamefont {Nesseris}\ \emph {et~al.}(2025)\citenamefont {Nesseris}, \citenamefont {Akrami},\ and\ \citenamefont {Starkman}}]{Nesseris:2025lke}%
  \BibitemOpen
  \bibfield  {author} {\bibinfo {author} {\bibfnamefont {Savvas}\ \bibnamefont {Nesseris}}, \bibinfo {author} {\bibfnamefont {Yashar}\ \bibnamefont {Akrami}}, \ and\ \bibinfo {author} {\bibfnamefont {Glenn~D.}\ \bibnamefont {Starkman}},\ }\bibfield  {title} {\enquote {\bibinfo {title} {{To CPL, or not to CPL? What we have not learned about the dark energy equation of state}},}\ }\href@noop {} {\  (\bibinfo {year} {2025})},\ \Eprint {http://arxiv.org/abs/2503.22529} {arXiv:2503.22529 [astro-ph.CO]} \BibitemShut {NoStop}%
\bibitem [{\citenamefont {Huterer}\ and\ \citenamefont {Starkman}(2003)}]{Huterer:2002hy}%
  \BibitemOpen
  \bibfield  {author} {\bibinfo {author} {\bibfnamefont {Dragan}\ \bibnamefont {Huterer}}\ and\ \bibinfo {author} {\bibfnamefont {Glenn}\ \bibnamefont {Starkman}},\ }\bibfield  {title} {\enquote {\bibinfo {title} {{Parameterization of dark-energy properties: A Principal-component approach}},}\ }\href {\doibase 10.1103/PhysRevLett.90.031301} {\bibfield  {journal} {\bibinfo  {journal} {Phys. Rev. Lett.}\ }\textbf {\bibinfo {volume} {90}},\ \bibinfo {pages} {031301} (\bibinfo {year} {2003})},\ \Eprint {http://arxiv.org/abs/astro-ph/0207517} {arXiv:astro-ph/0207517} \BibitemShut {NoStop}%
\bibitem [{\citenamefont {Agrawal}\ \emph {et~al.}(2018)\citenamefont {Agrawal}, \citenamefont {Obied}, \citenamefont {Steinhardt},\ and\ \citenamefont {Vafa}}]{Agrawal:2018own}%
  \BibitemOpen
  \bibfield  {author} {\bibinfo {author} {\bibfnamefont {Prateek}\ \bibnamefont {Agrawal}}, \bibinfo {author} {\bibfnamefont {Georges}\ \bibnamefont {Obied}}, \bibinfo {author} {\bibfnamefont {Paul~J.}\ \bibnamefont {Steinhardt}}, \ and\ \bibinfo {author} {\bibfnamefont {Cumrun}\ \bibnamefont {Vafa}},\ }\bibfield  {title} {\enquote {\bibinfo {title} {{On the Cosmological Implications of the String Swampland}},}\ }\href {\doibase 10.1016/j.physletb.2018.07.040} {\bibfield  {journal} {\bibinfo  {journal} {Phys. Lett. B}\ }\textbf {\bibinfo {volume} {784}},\ \bibinfo {pages} {271--276} (\bibinfo {year} {2018})},\ \Eprint {http://arxiv.org/abs/1806.09718} {arXiv:1806.09718 [hep-th]} \BibitemShut {NoStop}%
\bibitem [{\citenamefont {Cicoli}\ \emph {et~al.}(2019)\citenamefont {Cicoli}, \citenamefont {De~Alwis}, \citenamefont {Maharana}, \citenamefont {Muia},\ and\ \citenamefont {Quevedo}}]{Cicoli:2018kdo}%
  \BibitemOpen
  \bibfield  {author} {\bibinfo {author} {\bibfnamefont {Michele}\ \bibnamefont {Cicoli}}, \bibinfo {author} {\bibfnamefont {Senarath}\ \bibnamefont {De~Alwis}}, \bibinfo {author} {\bibfnamefont {Anshuman}\ \bibnamefont {Maharana}}, \bibinfo {author} {\bibfnamefont {Francesco}\ \bibnamefont {Muia}}, \ and\ \bibinfo {author} {\bibfnamefont {Fernando}\ \bibnamefont {Quevedo}},\ }\bibfield  {title} {\enquote {\bibinfo {title} {{De Sitter vs Quintessence in String Theory}},}\ }\href {\doibase 10.1002/prop.201800079} {\bibfield  {journal} {\bibinfo  {journal} {Fortsch. Phys.}\ }\textbf {\bibinfo {volume} {67}},\ \bibinfo {pages} {1800079} (\bibinfo {year} {2019})},\ \Eprint {http://arxiv.org/abs/1808.08967} {arXiv:1808.08967 [hep-th]} \BibitemShut {NoStop}%
\bibitem [{\citenamefont {Akrami}\ \emph {et~al.}(2019)\citenamefont {Akrami}, \citenamefont {Kallosh}, \citenamefont {Linde},\ and\ \citenamefont {Vardanyan}}]{Akrami:2018ylq}%
  \BibitemOpen
  \bibfield  {author} {\bibinfo {author} {\bibfnamefont {Yashar}\ \bibnamefont {Akrami}}, \bibinfo {author} {\bibfnamefont {Renata}\ \bibnamefont {Kallosh}}, \bibinfo {author} {\bibfnamefont {Andrei}\ \bibnamefont {Linde}}, \ and\ \bibinfo {author} {\bibfnamefont {Valeri}\ \bibnamefont {Vardanyan}},\ }\bibfield  {title} {\enquote {\bibinfo {title} {{The Landscape, the Swampland and the Era of Precision Cosmology}},}\ }\href {\doibase 10.1002/prop.201800075} {\bibfield  {journal} {\bibinfo  {journal} {Fortsch. Phys.}\ }\textbf {\bibinfo {volume} {67}},\ \bibinfo {pages} {1800075} (\bibinfo {year} {2019})},\ \Eprint {http://arxiv.org/abs/1808.09440} {arXiv:1808.09440 [hep-th]} \BibitemShut {NoStop}%
\bibitem [{\citenamefont {Han}\ \emph {et~al.}(2019)\citenamefont {Han}, \citenamefont {Pi},\ and\ \citenamefont {Sasaki}}]{Han:2018yrk}%
  \BibitemOpen
  \bibfield  {author} {\bibinfo {author} {\bibfnamefont {Chengcheng}\ \bibnamefont {Han}}, \bibinfo {author} {\bibfnamefont {Shi}\ \bibnamefont {Pi}}, \ and\ \bibinfo {author} {\bibfnamefont {Misao}\ \bibnamefont {Sasaki}},\ }\bibfield  {title} {\enquote {\bibinfo {title} {{Quintessence Saves Higgs Instability}},}\ }\href {\doibase 10.1016/j.physletb.2019.02.037} {\bibfield  {journal} {\bibinfo  {journal} {Phys. Lett. B}\ }\textbf {\bibinfo {volume} {791}},\ \bibinfo {pages} {314--318} (\bibinfo {year} {2019})},\ \Eprint {http://arxiv.org/abs/1809.05507} {arXiv:1809.05507 [hep-ph]} \BibitemShut {NoStop}%
\bibitem [{\citenamefont {Raveri}\ \emph {et~al.}(2019)\citenamefont {Raveri}, \citenamefont {Hu},\ and\ \citenamefont {Sethi}}]{Raveri:2018ddi}%
  \BibitemOpen
  \bibfield  {author} {\bibinfo {author} {\bibfnamefont {Marco}\ \bibnamefont {Raveri}}, \bibinfo {author} {\bibfnamefont {Wayne}\ \bibnamefont {Hu}}, \ and\ \bibinfo {author} {\bibfnamefont {Savdeep}\ \bibnamefont {Sethi}},\ }\bibfield  {title} {\enquote {\bibinfo {title} {{Swampland Conjectures and Late-Time Cosmology}},}\ }\href {\doibase 10.1103/PhysRevD.99.083518} {\bibfield  {journal} {\bibinfo  {journal} {Phys. Rev. D}\ }\textbf {\bibinfo {volume} {99}},\ \bibinfo {pages} {083518} (\bibinfo {year} {2019})},\ \Eprint {http://arxiv.org/abs/1812.10448} {arXiv:1812.10448 [hep-th]} \BibitemShut {NoStop}%
\bibitem [{\citenamefont {Akrami}\ \emph {et~al.}(2021{\natexlab{b}})\citenamefont {Akrami}, \citenamefont {Sasaki}, \citenamefont {Solomon},\ and\ \citenamefont {Vardanyan}}]{Akrami:2020zfz}%
  \BibitemOpen
  \bibfield  {author} {\bibinfo {author} {\bibfnamefont {Yashar}\ \bibnamefont {Akrami}}, \bibinfo {author} {\bibfnamefont {Misao}\ \bibnamefont {Sasaki}}, \bibinfo {author} {\bibfnamefont {Adam~R.}\ \bibnamefont {Solomon}}, \ and\ \bibinfo {author} {\bibfnamefont {Valeri}\ \bibnamefont {Vardanyan}},\ }\bibfield  {title} {\enquote {\bibinfo {title} {{Multi-field dark energy: Cosmic acceleration on a steep potential}},}\ }\href {\doibase 10.1016/j.physletb.2021.136427} {\bibfield  {journal} {\bibinfo  {journal} {Phys. Lett. B}\ }\textbf {\bibinfo {volume} {819}},\ \bibinfo {pages} {136427} (\bibinfo {year} {2021}{\natexlab{b}})},\ \Eprint {http://arxiv.org/abs/2008.13660} {arXiv:2008.13660 [astro-ph.CO]} \BibitemShut {NoStop}%
\bibitem [{\citenamefont {Andriot}\ \emph {et~al.}(2024)\citenamefont {Andriot}, \citenamefont {Parameswaran}, \citenamefont {Tsimpis}, \citenamefont {Wrase},\ and\ \citenamefont {Zavala}}]{Andriot:2024jsh}%
  \BibitemOpen
  \bibfield  {author} {\bibinfo {author} {\bibfnamefont {David}\ \bibnamefont {Andriot}}, \bibinfo {author} {\bibfnamefont {Susha}\ \bibnamefont {Parameswaran}}, \bibinfo {author} {\bibfnamefont {Dimitrios}\ \bibnamefont {Tsimpis}}, \bibinfo {author} {\bibfnamefont {Timm}\ \bibnamefont {Wrase}}, \ and\ \bibinfo {author} {\bibfnamefont {Ivonne}\ \bibnamefont {Zavala}},\ }\bibfield  {title} {\enquote {\bibinfo {title} {{Exponential quintessence: curved, steep and stringy?}}}\ }\href {\doibase 10.1007/JHEP08(2024)117} {\bibfield  {journal} {\bibinfo  {journal} {JHEP}\ }\textbf {\bibinfo {volume} {08}},\ \bibinfo {pages} {117} (\bibinfo {year} {2024})},\ \Eprint {http://arxiv.org/abs/2405.09323} {arXiv:2405.09323 [hep-th]} \BibitemShut {NoStop}%
\bibitem [{\citenamefont {Bhattacharya}\ \emph {et~al.}(2024{\natexlab{a}})\citenamefont {Bhattacharya}, \citenamefont {Borghetto}, \citenamefont {Malhotra}, \citenamefont {Parameswaran}, \citenamefont {Tasinato},\ and\ \citenamefont {Zavala}}]{Bhattacharya:2024hep}%
  \BibitemOpen
  \bibfield  {author} {\bibinfo {author} {\bibfnamefont {Sukannya}\ \bibnamefont {Bhattacharya}}, \bibinfo {author} {\bibfnamefont {Giulia}\ \bibnamefont {Borghetto}}, \bibinfo {author} {\bibfnamefont {Ameek}\ \bibnamefont {Malhotra}}, \bibinfo {author} {\bibfnamefont {Susha}\ \bibnamefont {Parameswaran}}, \bibinfo {author} {\bibfnamefont {Gianmassimo}\ \bibnamefont {Tasinato}}, \ and\ \bibinfo {author} {\bibfnamefont {Ivonne}\ \bibnamefont {Zavala}},\ }\bibfield  {title} {\enquote {\bibinfo {title} {{Cosmological constraints on curved quintessence}},}\ }\href {\doibase 10.1088/1475-7516/2024/09/073} {\bibfield  {journal} {\bibinfo  {journal} {JCAP}\ }\textbf {\bibinfo {volume} {09}},\ \bibinfo {pages} {073} (\bibinfo {year} {2024}{\natexlab{a}})},\ \Eprint {http://arxiv.org/abs/2405.17396} {arXiv:2405.17396 [astro-ph.CO]} \BibitemShut {NoStop}%
\bibitem [{\citenamefont {Andriot}(2024)}]{Andriot:2024sif}%
  \BibitemOpen
  \bibfield  {author} {\bibinfo {author} {\bibfnamefont {David}\ \bibnamefont {Andriot}},\ }\bibfield  {title} {\enquote {\bibinfo {title} {{Quintessence: an analytical study, with theoretical and observational applications}},}\ }\href@noop {} {\  (\bibinfo {year} {2024})},\ \Eprint {http://arxiv.org/abs/2410.17182} {arXiv:2410.17182 [hep-th]} \BibitemShut {NoStop}%
\bibitem [{\citenamefont {Alestas}\ \emph {et~al.}(2024)\citenamefont {Alestas}, \citenamefont {Delgado}, \citenamefont {Ruiz}, \citenamefont {Akrami}, \citenamefont {Montero},\ and\ \citenamefont {Nesseris}}]{Alestas:2024gxe}%
  \BibitemOpen
  \bibfield  {author} {\bibinfo {author} {\bibfnamefont {George}\ \bibnamefont {Alestas}}, \bibinfo {author} {\bibfnamefont {Matilda}\ \bibnamefont {Delgado}}, \bibinfo {author} {\bibfnamefont {Ignacio}\ \bibnamefont {Ruiz}}, \bibinfo {author} {\bibfnamefont {Yashar}\ \bibnamefont {Akrami}}, \bibinfo {author} {\bibfnamefont {Miguel}\ \bibnamefont {Montero}}, \ and\ \bibinfo {author} {\bibfnamefont {Savvas}\ \bibnamefont {Nesseris}},\ }\bibfield  {title} {\enquote {\bibinfo {title} {{Is curvature-assisted quintessence observationally viable?}}}\ }\href {\doibase 10.1103/PhysRevD.110.106010} {\bibfield  {journal} {\bibinfo  {journal} {Phys. Rev. D}\ }\textbf {\bibinfo {volume} {110}},\ \bibinfo {pages} {106010} (\bibinfo {year} {2024})},\ \Eprint {http://arxiv.org/abs/2406.09212} {arXiv:2406.09212 [hep-th]} \BibitemShut {NoStop}%
\bibitem [{\citenamefont {Bhattacharya}\ \emph {et~al.}(2024{\natexlab{b}})\citenamefont {Bhattacharya}, \citenamefont {Borghetto}, \citenamefont {Malhotra}, \citenamefont {Parameswaran}, \citenamefont {Tasinato},\ and\ \citenamefont {Zavala}}]{Bhattacharya:2024kxp}%
  \BibitemOpen
  \bibfield  {author} {\bibinfo {author} {\bibfnamefont {Sukannya}\ \bibnamefont {Bhattacharya}}, \bibinfo {author} {\bibfnamefont {Giulia}\ \bibnamefont {Borghetto}}, \bibinfo {author} {\bibfnamefont {Ameek}\ \bibnamefont {Malhotra}}, \bibinfo {author} {\bibfnamefont {Susha}\ \bibnamefont {Parameswaran}}, \bibinfo {author} {\bibfnamefont {Gianmassimo}\ \bibnamefont {Tasinato}}, \ and\ \bibinfo {author} {\bibfnamefont {Ivonne}\ \bibnamefont {Zavala}},\ }\bibfield  {title} {\enquote {\bibinfo {title} {{Cosmological tests of quintessence in quantum gravity}},}\ }\href@noop {} {\  (\bibinfo {year} {2024}{\natexlab{b}})},\ \Eprint {http://arxiv.org/abs/2410.21243} {arXiv:2410.21243 [astro-ph.CO]} \BibitemShut {NoStop}%
\bibitem [{\citenamefont {Andriot}\ \emph {et~al.}(2025)\citenamefont {Andriot}, \citenamefont {Rajaguru},\ and\ \citenamefont {Tringas}}]{Andriot:2025gyr}%
  \BibitemOpen
  \bibfield  {author} {\bibinfo {author} {\bibfnamefont {David}\ \bibnamefont {Andriot}}, \bibinfo {author} {\bibfnamefont {Muthusamy}\ \bibnamefont {Rajaguru}}, \ and\ \bibinfo {author} {\bibfnamefont {George}\ \bibnamefont {Tringas}},\ }\bibfield  {title} {\enquote {\bibinfo {title} {{Single versus multifield scalar potentials from string theory}},}\ }\href@noop {} {\  (\bibinfo {year} {2025})},\ \Eprint {http://arxiv.org/abs/2501.17775} {arXiv:2501.17775 [hep-th]} \BibitemShut {NoStop}%
\bibitem [{\citenamefont {Maldacena}\ and\ \citenamefont {Nunez}(2001)}]{Maldacena:2000mw}%
  \BibitemOpen
  \bibfield  {author} {\bibinfo {author} {\bibfnamefont {Juan~Martin}\ \bibnamefont {Maldacena}}\ and\ \bibinfo {author} {\bibfnamefont {Carlos}\ \bibnamefont {Nunez}},\ }\bibfield  {title} {\enquote {\bibinfo {title} {{Supergravity description of field theories on curved manifolds and a no go theorem}},}\ }\href {\doibase 10.1142/S0217751X01003937} {\bibfield  {journal} {\bibinfo  {journal} {Int. J. Mod. Phys. A}\ }\textbf {\bibinfo {volume} {16}},\ \bibinfo {pages} {822--855} (\bibinfo {year} {2001})},\ \Eprint {http://arxiv.org/abs/hep-th/0007018} {arXiv:hep-th/0007018} \BibitemShut {NoStop}%
\bibitem [{\citenamefont {Hertzberg}\ \emph {et~al.}(2007)\citenamefont {Hertzberg}, \citenamefont {Kachru}, \citenamefont {Taylor},\ and\ \citenamefont {Tegmark}}]{Hertzberg:2007wc}%
  \BibitemOpen
  \bibfield  {author} {\bibinfo {author} {\bibfnamefont {Mark~P.}\ \bibnamefont {Hertzberg}}, \bibinfo {author} {\bibfnamefont {Shamit}\ \bibnamefont {Kachru}}, \bibinfo {author} {\bibfnamefont {Washington}\ \bibnamefont {Taylor}}, \ and\ \bibinfo {author} {\bibfnamefont {Max}\ \bibnamefont {Tegmark}},\ }\bibfield  {title} {\enquote {\bibinfo {title} {{Inflationary Constraints on Type IIA String Theory}},}\ }\href {\doibase 10.1088/1126-6708/2007/12/095} {\bibfield  {journal} {\bibinfo  {journal} {JHEP}\ }\textbf {\bibinfo {volume} {12}},\ \bibinfo {pages} {095} (\bibinfo {year} {2007})},\ \Eprint {http://arxiv.org/abs/0711.2512} {arXiv:0711.2512 [hep-th]} \BibitemShut {NoStop}%
\bibitem [{\citenamefont {Obied}\ \emph {et~al.}(2018)\citenamefont {Obied}, \citenamefont {Ooguri}, \citenamefont {Spodyneiko},\ and\ \citenamefont {Vafa}}]{Obied:2018sgi}%
  \BibitemOpen
  \bibfield  {author} {\bibinfo {author} {\bibfnamefont {Georges}\ \bibnamefont {Obied}}, \bibinfo {author} {\bibfnamefont {Hirosi}\ \bibnamefont {Ooguri}}, \bibinfo {author} {\bibfnamefont {Lev}\ \bibnamefont {Spodyneiko}}, \ and\ \bibinfo {author} {\bibfnamefont {Cumrun}\ \bibnamefont {Vafa}},\ }\bibfield  {title} {\enquote {\bibinfo {title} {{De Sitter Space and the Swampland}},}\ }\href@noop {} {\  (\bibinfo {year} {2018})},\ \Eprint {http://arxiv.org/abs/1806.08362} {arXiv:1806.08362 [hep-th]} \BibitemShut {NoStop}%
\bibitem [{\citenamefont {Andriot}(2019)}]{Andriot:2019wrs}%
  \BibitemOpen
  \bibfield  {author} {\bibinfo {author} {\bibfnamefont {David}\ \bibnamefont {Andriot}},\ }\bibfield  {title} {\enquote {\bibinfo {title} {{Open problems on classical de Sitter solutions}},}\ }\href {\doibase 10.1002/prop.201900026} {\bibfield  {journal} {\bibinfo  {journal} {Fortsch. Phys.}\ }\textbf {\bibinfo {volume} {67}},\ \bibinfo {pages} {1900026} (\bibinfo {year} {2019})},\ \Eprint {http://arxiv.org/abs/1902.10093} {arXiv:1902.10093 [hep-th]} \BibitemShut {NoStop}%
\bibitem [{\citenamefont {Andriot}\ \emph {et~al.}(2020)\citenamefont {Andriot}, \citenamefont {Cribiori},\ and\ \citenamefont {Erkinger}}]{Andriot:2020lea}%
  \BibitemOpen
  \bibfield  {author} {\bibinfo {author} {\bibfnamefont {David}\ \bibnamefont {Andriot}}, \bibinfo {author} {\bibfnamefont {Niccol\`o}\ \bibnamefont {Cribiori}}, \ and\ \bibinfo {author} {\bibfnamefont {David}\ \bibnamefont {Erkinger}},\ }\bibfield  {title} {\enquote {\bibinfo {title} {{The web of swampland conjectures and the TCC bound}},}\ }\href {\doibase 10.1007/JHEP07(2020)162} {\bibfield  {journal} {\bibinfo  {journal} {JHEP}\ }\textbf {\bibinfo {volume} {07}},\ \bibinfo {pages} {162} (\bibinfo {year} {2020})},\ \Eprint {http://arxiv.org/abs/2004.00030} {arXiv:2004.00030 [hep-th]} \BibitemShut {NoStop}%
\bibitem [{\citenamefont {Calder\'on-Infante}\ \emph {et~al.}(2023)\citenamefont {Calder\'on-Infante}, \citenamefont {Ruiz},\ and\ \citenamefont {Valenzuela}}]{Calderon-Infante:2022nxb}%
  \BibitemOpen
  \bibfield  {author} {\bibinfo {author} {\bibfnamefont {Jos\'e}\ \bibnamefont {Calder\'on-Infante}}, \bibinfo {author} {\bibfnamefont {Ignacio}\ \bibnamefont {Ruiz}}, \ and\ \bibinfo {author} {\bibfnamefont {Irene}\ \bibnamefont {Valenzuela}},\ }\bibfield  {title} {\enquote {\bibinfo {title} {{Asymptotic accelerated expansion in string theory and the Swampland}},}\ }\href {\doibase 10.1007/JHEP06(2023)129} {\bibfield  {journal} {\bibinfo  {journal} {JHEP}\ }\textbf {\bibinfo {volume} {06}},\ \bibinfo {pages} {129} (\bibinfo {year} {2023})},\ \Eprint {http://arxiv.org/abs/2209.11821} {arXiv:2209.11821 [hep-th]} \BibitemShut {NoStop}%
\bibitem [{\citenamefont {Shiu}\ \emph {et~al.}(2023{\natexlab{a}})\citenamefont {Shiu}, \citenamefont {Tonioni},\ and\ \citenamefont {Tran}}]{Shiu:2023fhb}%
  \BibitemOpen
  \bibfield  {author} {\bibinfo {author} {\bibfnamefont {Gary}\ \bibnamefont {Shiu}}, \bibinfo {author} {\bibfnamefont {Flavio}\ \bibnamefont {Tonioni}}, \ and\ \bibinfo {author} {\bibfnamefont {Hung~V.}\ \bibnamefont {Tran}},\ }\bibfield  {title} {\enquote {\bibinfo {title} {{Late-time attractors and cosmic acceleration}},}\ }\href {\doibase 10.1103/PhysRevD.108.063528} {\bibfield  {journal} {\bibinfo  {journal} {Phys. Rev. D}\ }\textbf {\bibinfo {volume} {108}},\ \bibinfo {pages} {063528} (\bibinfo {year} {2023}{\natexlab{a}})},\ \Eprint {http://arxiv.org/abs/2306.07327} {arXiv:2306.07327 [hep-th]} \BibitemShut {NoStop}%
\bibitem [{\citenamefont {Shiu}\ \emph {et~al.}(2023{\natexlab{b}})\citenamefont {Shiu}, \citenamefont {Tonioni},\ and\ \citenamefont {Tran}}]{Shiu:2023nph}%
  \BibitemOpen
  \bibfield  {author} {\bibinfo {author} {\bibfnamefont {Gary}\ \bibnamefont {Shiu}}, \bibinfo {author} {\bibfnamefont {Flavio}\ \bibnamefont {Tonioni}}, \ and\ \bibinfo {author} {\bibfnamefont {Hung~V.}\ \bibnamefont {Tran}},\ }\bibfield  {title} {\enquote {\bibinfo {title} {{Accelerating universe at the end of time}},}\ }\href {\doibase 10.1103/PhysRevD.108.063527} {\bibfield  {journal} {\bibinfo  {journal} {Phys. Rev. D}\ }\textbf {\bibinfo {volume} {108}},\ \bibinfo {pages} {063527} (\bibinfo {year} {2023}{\natexlab{b}})},\ \Eprint {http://arxiv.org/abs/2303.03418} {arXiv:2303.03418 [hep-th]} \BibitemShut {NoStop}%
\bibitem [{\citenamefont {Cremonini}\ \emph {et~al.}(2023)\citenamefont {Cremonini}, \citenamefont {Gonzalo}, \citenamefont {Rajaguru}, \citenamefont {Tang},\ and\ \citenamefont {Wrase}}]{Cremonini:2023suw}%
  \BibitemOpen
  \bibfield  {author} {\bibinfo {author} {\bibfnamefont {Sera}\ \bibnamefont {Cremonini}}, \bibinfo {author} {\bibfnamefont {Eduardo}\ \bibnamefont {Gonzalo}}, \bibinfo {author} {\bibfnamefont {Muthusamy}\ \bibnamefont {Rajaguru}}, \bibinfo {author} {\bibfnamefont {Yuezhang}\ \bibnamefont {Tang}}, \ and\ \bibinfo {author} {\bibfnamefont {Timm}\ \bibnamefont {Wrase}},\ }\bibfield  {title} {\enquote {\bibinfo {title} {{On asymptotic dark energy in string theory}},}\ }\href {\doibase 10.1007/JHEP09(2023)075} {\bibfield  {journal} {\bibinfo  {journal} {JHEP}\ }\textbf {\bibinfo {volume} {09}},\ \bibinfo {pages} {075} (\bibinfo {year} {2023})},\ \Eprint {http://arxiv.org/abs/2306.15714} {arXiv:2306.15714 [hep-th]} \BibitemShut {NoStop}%
\bibitem [{\citenamefont {Hebecker}\ \emph {et~al.}(2023)\citenamefont {Hebecker}, \citenamefont {Schreyer},\ and\ \citenamefont {Venken}}]{Hebecker:2023qke}%
  \BibitemOpen
  \bibfield  {author} {\bibinfo {author} {\bibfnamefont {Arthur}\ \bibnamefont {Hebecker}}, \bibinfo {author} {\bibfnamefont {Simon}\ \bibnamefont {Schreyer}}, \ and\ \bibinfo {author} {\bibfnamefont {Gerben}\ \bibnamefont {Venken}},\ }\bibfield  {title} {\enquote {\bibinfo {title} {{No asymptotic acceleration without higher-dimensional de Sitter vacua}},}\ }\href {\doibase 10.1007/JHEP11(2023)173} {\bibfield  {journal} {\bibinfo  {journal} {JHEP}\ }\textbf {\bibinfo {volume} {11}},\ \bibinfo {pages} {173} (\bibinfo {year} {2023})},\ \Eprint {http://arxiv.org/abs/2306.17213} {arXiv:2306.17213 [hep-th]} \BibitemShut {NoStop}%
\bibitem [{\citenamefont {Van~Riet}(2024)}]{VanRiet:2023cca}%
  \BibitemOpen
  \bibfield  {author} {\bibinfo {author} {\bibfnamefont {Thomas}\ \bibnamefont {Van~Riet}},\ }\bibfield  {title} {\enquote {\bibinfo {title} {{No accelerating scaling cosmologies at string tree level?}}}\ }\href {\doibase 10.1088/1475-7516/2024/01/055} {\bibfield  {journal} {\bibinfo  {journal} {JCAP}\ }\textbf {\bibinfo {volume} {01}},\ \bibinfo {pages} {055} (\bibinfo {year} {2024})},\ \Eprint {http://arxiv.org/abs/2308.15035} {arXiv:2308.15035 [hep-th]} \BibitemShut {NoStop}%
\bibitem [{\citenamefont {Seo}(2024)}]{Seo:2024fki}%
  \BibitemOpen
  \bibfield  {author} {\bibinfo {author} {\bibfnamefont {Min-Seok}\ \bibnamefont {Seo}},\ }\bibfield  {title} {\enquote {\bibinfo {title} {{Asymptotic bound on slow-roll parameter in stringy quintessence model}},}\ }\href@noop {} {\bibfield  {journal} {\bibinfo  {journal} {JCAP}\ } (\bibinfo {year} {2024})},\ \Eprint {http://arxiv.org/abs/2402.00241} {arXiv:2402.00241 [hep-th]} \BibitemShut {NoStop}%
\bibitem [{\citenamefont {Rudelius}(2021)}]{Rudelius:2021azq}%
  \BibitemOpen
  \bibfield  {author} {\bibinfo {author} {\bibfnamefont {Tom}\ \bibnamefont {Rudelius}},\ }\bibfield  {title} {\enquote {\bibinfo {title} {{Asymptotic observables and the swampland}},}\ }\href {\doibase 10.1103/PhysRevD.104.126023} {\bibfield  {journal} {\bibinfo  {journal} {Phys. Rev. D}\ }\textbf {\bibinfo {volume} {104}},\ \bibinfo {pages} {126023} (\bibinfo {year} {2021})},\ \Eprint {http://arxiv.org/abs/2106.09026} {arXiv:2106.09026 [hep-th]} \BibitemShut {NoStop}%
\bibitem [{\citenamefont {Ramadan}\ \emph {et~al.}(2024)\citenamefont {Ramadan}, \citenamefont {Sakstein},\ and\ \citenamefont {Rubin}}]{Ramadan:2024kmn}%
  \BibitemOpen
  \bibfield  {author} {\bibinfo {author} {\bibfnamefont {Omar~F.}\ \bibnamefont {Ramadan}}, \bibinfo {author} {\bibfnamefont {Jeremy}\ \bibnamefont {Sakstein}}, \ and\ \bibinfo {author} {\bibfnamefont {David}\ \bibnamefont {Rubin}},\ }\bibfield  {title} {\enquote {\bibinfo {title} {{DESI constraints on exponential quintessence}},}\ }\href {\doibase 10.1103/PhysRevD.110.L041303} {\bibfield  {journal} {\bibinfo  {journal} {Phys. Rev. D}\ }\textbf {\bibinfo {volume} {110}},\ \bibinfo {pages} {L041303} (\bibinfo {year} {2024})},\ \Eprint {http://arxiv.org/abs/2405.18747} {arXiv:2405.18747 [astro-ph.CO]} \BibitemShut {NoStop}%
\bibitem [{\citenamefont {Aghanim}\ \emph {et~al.}(2020)\citenamefont {Aghanim} \emph {et~al.}}]{Planck:2018vyg}%
  \BibitemOpen
  \bibfield  {author} {\bibinfo {author} {\bibfnamefont {N.}~\bibnamefont {Aghanim}} \emph {et~al.} (\bibinfo {collaboration} {Planck}),\ }\bibfield  {title} {\enquote {\bibinfo {title} {{Planck 2018 results. VI. Cosmological parameters}},}\ }\href {\doibase 10.1051/0004-6361/201833910} {\bibfield  {journal} {\bibinfo  {journal} {Astron. Astrophys.}\ }\textbf {\bibinfo {volume} {641}},\ \bibinfo {pages} {A6} (\bibinfo {year} {2020})},\ \bibinfo {note} {[Erratum: Astron.Astrophys. 652, C4 (2021)]},\ \Eprint {http://arxiv.org/abs/1807.06209} {arXiv:1807.06209 [astro-ph.CO]} \BibitemShut {NoStop}%
\bibitem [{\citenamefont {Abbott}\ \emph {et~al.}(2024)\citenamefont {Abbott} \emph {et~al.}}]{DES:2024jxu}%
  \BibitemOpen
  \bibfield  {author} {\bibinfo {author} {\bibfnamefont {T.~M.~C.}\ \bibnamefont {Abbott}} \emph {et~al.} (\bibinfo {collaboration} {DES}),\ }\bibfield  {title} {\enquote {\bibinfo {title} {{The Dark Energy Survey: Cosmology Results with \ensuremath{\sim}1500 New High-redshift Type Ia Supernovae Using the Full 5 yr Data Set}},}\ }\href {\doibase 10.3847/2041-8213/ad6f9f} {\bibfield  {journal} {\bibinfo  {journal} {Astrophys. J. Lett.}\ }\textbf {\bibinfo {volume} {973}},\ \bibinfo {pages} {L14} (\bibinfo {year} {2024})},\ \Eprint {http://arxiv.org/abs/2401.02929} {arXiv:2401.02929 [astro-ph.CO]} \BibitemShut {NoStop}%
\bibitem [{\citenamefont {Zhai}\ and\ \citenamefont {Wang}(2019)}]{Zhai:2018vmm}%
  \BibitemOpen
  \bibfield  {author} {\bibinfo {author} {\bibfnamefont {Zhongxu}\ \bibnamefont {Zhai}}\ and\ \bibinfo {author} {\bibfnamefont {Yun}\ \bibnamefont {Wang}},\ }\bibfield  {title} {\enquote {\bibinfo {title} {{Robust and model-independent cosmological constraints from distance measurements}},}\ }\href {\doibase 10.1088/1475-7516/2019/07/005} {\bibfield  {journal} {\bibinfo  {journal} {JCAP}\ }\textbf {\bibinfo {volume} {07}},\ \bibinfo {pages} {005} (\bibinfo {year} {2019})},\ \Eprint {http://arxiv.org/abs/1811.07425} {arXiv:1811.07425 [astro-ph.CO]} \BibitemShut {NoStop}%
\bibitem [{\citenamefont {Brieden}\ \emph {et~al.}(2023)\citenamefont {Brieden}, \citenamefont {Gil-Mar\'\i{}n},\ and\ \citenamefont {Verde}}]{Brieden:2022heh}%
  \BibitemOpen
  \bibfield  {author} {\bibinfo {author} {\bibfnamefont {Samuel}\ \bibnamefont {Brieden}}, \bibinfo {author} {\bibfnamefont {H\'ector}\ \bibnamefont {Gil-Mar\'\i{}n}}, \ and\ \bibinfo {author} {\bibfnamefont {Licia}\ \bibnamefont {Verde}},\ }\bibfield  {title} {\enquote {\bibinfo {title} {{A tale of two (or more) h's}},}\ }\href {\doibase 10.1088/1475-7516/2023/04/023} {\bibfield  {journal} {\bibinfo  {journal} {JCAP}\ }\textbf {\bibinfo {volume} {04}},\ \bibinfo {pages} {023} (\bibinfo {year} {2023})},\ \Eprint {http://arxiv.org/abs/2212.04522} {arXiv:2212.04522 [astro-ph.CO]} \BibitemShut {NoStop}%
\bibitem [{\citenamefont {Sch\"oneberg}(2024)}]{Schoneberg:2024ifp}%
  \BibitemOpen
  \bibfield  {author} {\bibinfo {author} {\bibfnamefont {Nils}\ \bibnamefont {Sch\"oneberg}},\ }\bibfield  {title} {\enquote {\bibinfo {title} {{The 2024 BBN baryon abundance update}},}\ }\href {\doibase 10.1088/1475-7516/2024/06/006} {\bibfield  {journal} {\bibinfo  {journal} {JCAP}\ }\textbf {\bibinfo {volume} {06}},\ \bibinfo {pages} {006} (\bibinfo {year} {2024})},\ \Eprint {http://arxiv.org/abs/2401.15054} {arXiv:2401.15054 [astro-ph.CO]} \BibitemShut {NoStop}%
\bibitem [{\citenamefont {Heavens}\ \emph {et~al.}(2017)\citenamefont {Heavens}, \citenamefont {Fantaye}, \citenamefont {Mootoovaloo}, \citenamefont {Eggers}, \citenamefont {Hosenie}, \citenamefont {Kroon},\ and\ \citenamefont {Sellentin}}]{Heavens:2017afc}%
  \BibitemOpen
  \bibfield  {author} {\bibinfo {author} {\bibfnamefont {Alan}\ \bibnamefont {Heavens}}, \bibinfo {author} {\bibfnamefont {Yabebal}\ \bibnamefont {Fantaye}}, \bibinfo {author} {\bibfnamefont {Arrykrishna}\ \bibnamefont {Mootoovaloo}}, \bibinfo {author} {\bibfnamefont {Hans}\ \bibnamefont {Eggers}}, \bibinfo {author} {\bibfnamefont {Zafiirah}\ \bibnamefont {Hosenie}}, \bibinfo {author} {\bibfnamefont {Steve}\ \bibnamefont {Kroon}}, \ and\ \bibinfo {author} {\bibfnamefont {Elena}\ \bibnamefont {Sellentin}},\ }\bibfield  {title} {\enquote {\bibinfo {title} {{Marginal Likelihoods from Monte Carlo Markov Chains}},}\ }\href@noop {} {\  (\bibinfo {year} {2017})},\ \Eprint {http://arxiv.org/abs/1704.03472} {arXiv:1704.03472 [stat.CO]} \BibitemShut {NoStop}%
\bibitem [{\citenamefont {Jeffreys}(1939)}]{Jeffreys:1939xee}%
  \BibitemOpen
  \bibfield  {author} {\bibinfo {author} {\bibfnamefont {Harold}\ \bibnamefont {Jeffreys}},\ }\href@noop {} {\emph {\bibinfo {title} {{The Theory of Probability}}}},\ Oxford Classic Texts in the Physical Sciences\ (\bibinfo {year} {1939})\BibitemShut {NoStop}%
\bibitem [{\citenamefont {Trotta}(2008)}]{Trotta:2008qt}%
  \BibitemOpen
  \bibfield  {author} {\bibinfo {author} {\bibfnamefont {Roberto}\ \bibnamefont {Trotta}},\ }\bibfield  {title} {\enquote {\bibinfo {title} {{Bayes in the sky: Bayesian inference and model selection in cosmology}},}\ }\href {\doibase 10.1080/00107510802066753} {\bibfield  {journal} {\bibinfo  {journal} {Contemp. Phys.}\ }\textbf {\bibinfo {volume} {49}},\ \bibinfo {pages} {71--104} (\bibinfo {year} {2008})},\ \Eprint {http://arxiv.org/abs/0803.4089} {arXiv:0803.4089 [astro-ph]} \BibitemShut {NoStop}%
\bibitem [{\citenamefont {John}\ and\ \citenamefont {Narlikar}(2002)}]{John:2002gg}%
  \BibitemOpen
  \bibfield  {author} {\bibinfo {author} {\bibfnamefont {Moncy~V.}\ \bibnamefont {John}}\ and\ \bibinfo {author} {\bibfnamefont {J.~V.}\ \bibnamefont {Narlikar}},\ }\bibfield  {title} {\enquote {\bibinfo {title} {{Comparison of cosmological models using bayesian theory}},}\ }\href {\doibase 10.1103/PhysRevD.65.043506} {\bibfield  {journal} {\bibinfo  {journal} {Phys. Rev. D}\ }\textbf {\bibinfo {volume} {65}},\ \bibinfo {pages} {043506} (\bibinfo {year} {2002})},\ \Eprint {http://arxiv.org/abs/astro-ph/0111122} {arXiv:astro-ph/0111122} \BibitemShut {NoStop}%
\bibitem [{\citenamefont {Nesseris}\ and\ \citenamefont {Garcia-Bellido}(2013)}]{Nesseris:2012cq}%
  \BibitemOpen
  \bibfield  {author} {\bibinfo {author} {\bibfnamefont {Savvas}\ \bibnamefont {Nesseris}}\ and\ \bibinfo {author} {\bibfnamefont {Juan}\ \bibnamefont {Garcia-Bellido}},\ }\bibfield  {title} {\enquote {\bibinfo {title} {{Is the Jeffreys' scale a reliable tool for Bayesian model comparison in cosmology?}}}\ }\href {\doibase 10.1088/1475-7516/2013/08/036} {\bibfield  {journal} {\bibinfo  {journal} {JCAP}\ }\textbf {\bibinfo {volume} {08}},\ \bibinfo {pages} {036} (\bibinfo {year} {2013})},\ \Eprint {http://arxiv.org/abs/1210.7652} {arXiv:1210.7652 [astro-ph.CO]} \BibitemShut {NoStop}%
\bibitem [{\citenamefont {Nesseris}\ and\ \citenamefont {Perivolaropoulos}(2007)}]{Nesseris:2006er}%
  \BibitemOpen
  \bibfield  {author} {\bibinfo {author} {\bibfnamefont {S.}~\bibnamefont {Nesseris}}\ and\ \bibinfo {author} {\bibfnamefont {Leandros}\ \bibnamefont {Perivolaropoulos}},\ }\bibfield  {title} {\enquote {\bibinfo {title} {{Crossing the Phantom Divide: Theoretical Implications and Observational Status}},}\ }\href {\doibase 10.1088/1475-7516/2007/01/018} {\bibfield  {journal} {\bibinfo  {journal} {JCAP}\ }\textbf {\bibinfo {volume} {01}},\ \bibinfo {pages} {018} (\bibinfo {year} {2007})},\ \Eprint {http://arxiv.org/abs/astro-ph/0610092} {arXiv:astro-ph/0610092} \BibitemShut {NoStop}%
\bibitem [{\citenamefont {Sapone}\ and\ \citenamefont {Nesseris}(2024)}]{Sapone:2024ltl}%
  \BibitemOpen
  \bibfield  {author} {\bibinfo {author} {\bibfnamefont {Domenico}\ \bibnamefont {Sapone}}\ and\ \bibinfo {author} {\bibfnamefont {Savvas}\ \bibnamefont {Nesseris}},\ }\bibfield  {title} {\enquote {\bibinfo {title} {{Outliers in DESI BAO: robustness and cosmological implications}},}\ }\href@noop {} {\  (\bibinfo {year} {2024})},\ \Eprint {http://arxiv.org/abs/2412.01740} {arXiv:2412.01740 [astro-ph.CO]} \BibitemShut {NoStop}%
\bibitem [{\citenamefont {Colg\'ain}\ \emph {et~al.}(2024)\citenamefont {Colg\'ain}, \citenamefont {Dainotti}, \citenamefont {Capozziello}, \citenamefont {Pourojaghi}, \citenamefont {Sheikh-Jabbari},\ and\ \citenamefont {Stojkovic}}]{Colgain:2024xqj}%
  \BibitemOpen
  \bibfield  {author} {\bibinfo {author} {\bibfnamefont {Eoin~\'O.}\ \bibnamefont {Colg\'ain}}, \bibinfo {author} {\bibfnamefont {Maria~Giovanna}\ \bibnamefont {Dainotti}}, \bibinfo {author} {\bibfnamefont {Salvatore}\ \bibnamefont {Capozziello}}, \bibinfo {author} {\bibfnamefont {Saeed}\ \bibnamefont {Pourojaghi}}, \bibinfo {author} {\bibfnamefont {M.~M.}\ \bibnamefont {Sheikh-Jabbari}}, \ and\ \bibinfo {author} {\bibfnamefont {Dejan}\ \bibnamefont {Stojkovic}},\ }\bibfield  {title} {\enquote {\bibinfo {title} {{Does DESI 2024 Confirm $\Lambda$CDM?}}}\ }\href@noop {} {\  (\bibinfo {year} {2024})},\ \Eprint {http://arxiv.org/abs/2404.08633} {arXiv:2404.08633 [astro-ph.CO]} \BibitemShut {NoStop}%
\bibitem [{\citenamefont {Chudaykin}\ and\ \citenamefont {Kunz}(2024)}]{Chudaykin:2024gol}%
  \BibitemOpen
  \bibfield  {author} {\bibinfo {author} {\bibfnamefont {Anton}\ \bibnamefont {Chudaykin}}\ and\ \bibinfo {author} {\bibfnamefont {Martin}\ \bibnamefont {Kunz}},\ }\bibfield  {title} {\enquote {\bibinfo {title} {{Modified gravity interpretation of the evolving dark energy in light of DESI data}},}\ }\href {\doibase 10.1103/PhysRevD.110.123524} {\bibfield  {journal} {\bibinfo  {journal} {Phys. Rev. D}\ }\textbf {\bibinfo {volume} {110}},\ \bibinfo {pages} {123524} (\bibinfo {year} {2024})},\ \Eprint {http://arxiv.org/abs/2407.02558} {arXiv:2407.02558 [astro-ph.CO]} \BibitemShut {NoStop}%
\bibitem [{\citenamefont {Anselmi}\ \emph {et~al.}(2019)\citenamefont {Anselmi}, \citenamefont {Corasaniti}, \citenamefont {Sanchez}, \citenamefont {Starkman}, \citenamefont {Sheth},\ and\ \citenamefont {Zehavi}}]{Anselmi:2018vjz}%
  \BibitemOpen
  \bibfield  {author} {\bibinfo {author} {\bibfnamefont {Stefano}\ \bibnamefont {Anselmi}}, \bibinfo {author} {\bibfnamefont {Pier-Stefano}\ \bibnamefont {Corasaniti}}, \bibinfo {author} {\bibfnamefont {Ariel~G.}\ \bibnamefont {Sanchez}}, \bibinfo {author} {\bibfnamefont {Glenn~D.}\ \bibnamefont {Starkman}}, \bibinfo {author} {\bibfnamefont {Ravi~K.}\ \bibnamefont {Sheth}}, \ and\ \bibinfo {author} {\bibfnamefont {Idit}\ \bibnamefont {Zehavi}},\ }\bibfield  {title} {\enquote {\bibinfo {title} {{Cosmic distance inference from purely geometric BAO methods: Linear Point standard ruler and Correlation Function Model Fitting}},}\ }\href {\doibase 10.1103/PhysRevD.99.123515} {\bibfield  {journal} {\bibinfo  {journal} {Phys. Rev. D}\ }\textbf {\bibinfo {volume} {99}},\ \bibinfo {pages} {123515} (\bibinfo {year} {2019})},\ \Eprint {http://arxiv.org/abs/1811.12312} {arXiv:1811.12312 [astro-ph.CO]} \BibitemShut {NoStop}%
\bibitem [{\citenamefont {Anselmi}\ \emph {et~al.}(2023)\citenamefont {Anselmi}, \citenamefont {Starkman},\ and\ \citenamefont {Renzi}}]{Anselmi:2022exn}%
  \BibitemOpen
  \bibfield  {author} {\bibinfo {author} {\bibfnamefont {Stefano}\ \bibnamefont {Anselmi}}, \bibinfo {author} {\bibfnamefont {Glenn~D.}\ \bibnamefont {Starkman}}, \ and\ \bibinfo {author} {\bibfnamefont {Alessandro}\ \bibnamefont {Renzi}},\ }\bibfield  {title} {\enquote {\bibinfo {title} {{Cosmological forecasts for future galaxy surveys with the linear point standard ruler: Toward consistent BAO analyses far from a fiducial cosmology}},}\ }\href {\doibase 10.1103/PhysRevD.107.123506} {\bibfield  {journal} {\bibinfo  {journal} {Phys. Rev. D}\ }\textbf {\bibinfo {volume} {107}},\ \bibinfo {pages} {123506} (\bibinfo {year} {2023})},\ \Eprint {http://arxiv.org/abs/2205.09098} {arXiv:2205.09098 [astro-ph.CO]} \BibitemShut {NoStop}%
\bibitem [{\citenamefont {Nielsen}\ \emph {et~al.}(2016)\citenamefont {Nielsen}, \citenamefont {Guffanti},\ and\ \citenamefont {Sarkar}}]{Nielsen:2015pga}%
  \BibitemOpen
  \bibfield  {author} {\bibinfo {author} {\bibfnamefont {Jeppe~Tr\o{}st}\ \bibnamefont {Nielsen}}, \bibinfo {author} {\bibfnamefont {Alberto}\ \bibnamefont {Guffanti}}, \ and\ \bibinfo {author} {\bibfnamefont {Subir}\ \bibnamefont {Sarkar}},\ }\bibfield  {title} {\enquote {\bibinfo {title} {{Marginal evidence for cosmic acceleration from Type Ia supernovae}},}\ }\href {\doibase 10.1038/srep35596} {\bibfield  {journal} {\bibinfo  {journal} {Sci. Rep.}\ }\textbf {\bibinfo {volume} {6}},\ \bibinfo {pages} {35596} (\bibinfo {year} {2016})},\ \Eprint {http://arxiv.org/abs/1506.01354} {arXiv:1506.01354 [astro-ph.CO]} \BibitemShut {NoStop}%
\bibitem [{\citenamefont {Sah}\ \emph {et~al.}(2024)\citenamefont {Sah}, \citenamefont {Rameez}, \citenamefont {Sarkar},\ and\ \citenamefont {Tsagas}}]{Sah:2024csa}%
  \BibitemOpen
  \bibfield  {author} {\bibinfo {author} {\bibfnamefont {Animesh}\ \bibnamefont {Sah}}, \bibinfo {author} {\bibfnamefont {Mohamed}\ \bibnamefont {Rameez}}, \bibinfo {author} {\bibfnamefont {Subir}\ \bibnamefont {Sarkar}}, \ and\ \bibinfo {author} {\bibfnamefont {Christos}\ \bibnamefont {Tsagas}},\ }\bibfield  {title} {\enquote {\bibinfo {title} {{Anisotropy in Pantheon+ supernovae}},}\ }\href@noop {} {\  (\bibinfo {year} {2024})},\ \Eprint {http://arxiv.org/abs/2411.10838} {arXiv:2411.10838 [astro-ph.CO]} \BibitemShut {NoStop}%
\bibitem [{\citenamefont {{Akaike}}(1974)}]{AkaikeCrit}%
  \BibitemOpen
  \bibfield  {author} {\bibinfo {author} {\bibfnamefont {H.}~\bibnamefont {{Akaike}}},\ }\bibfield  {title} {\enquote {\bibinfo {title} {{A New Look at the Statistical Model Identification}},}\ }\href@noop {} {\bibfield  {journal} {\bibinfo  {journal} {IEEE Transactions on Automatic Control}\ }\textbf {\bibinfo {volume} {19}},\ \bibinfo {pages} {716--723} (\bibinfo {year} {1974})}\BibitemShut {NoStop}%
\bibitem [{\citenamefont {Schwarz}(1978)}]{Schwarz1978}%
  \BibitemOpen
  \bibfield  {author} {\bibinfo {author} {\bibfnamefont {Gideon}\ \bibnamefont {Schwarz}},\ }\bibfield  {title} {\enquote {\bibinfo {title} {Estimating the dimension of a model},}\ }\href {\doibase 10.1214/aos/1176344136} {\bibfield  {journal} {\bibinfo  {journal} {The Annals of Statistics}\ }\textbf {\bibinfo {volume} {6}},\ \bibinfo {pages} {461--464} (\bibinfo {year} {1978})}\BibitemShut {NoStop}%
\bibitem [{\citenamefont {Sellke}\ \emph {et~al.}(2001)\citenamefont {Sellke}, \citenamefont {Bayarri},\ and\ \citenamefont {and}}]{Sellke01022001}%
  \BibitemOpen
  \bibfield  {author} {\bibinfo {author} {\bibfnamefont {Thomas}\ \bibnamefont {Sellke}}, \bibinfo {author} {\bibfnamefont {M.~J}\ \bibnamefont {Bayarri}}, \ and\ \bibinfo {author} {\bibfnamefont {James O~Berger}\ \bibnamefont {and}},\ }\bibfield  {title} {\enquote {\bibinfo {title} {Calibration of p values for testing precise null hypotheses},}\ }\href {\doibase 10.1198/000313001300339950} {\bibfield  {journal} {\bibinfo  {journal} {The American Statistician}\ }\textbf {\bibinfo {volume} {55}},\ \bibinfo {pages} {62--71} (\bibinfo {year} {2001})}\BibitemShut {NoStop}%
\end{thebibliography}%
\end{document}